\documentclass[pre,twocolumn,amsmath,amssymb,nofootinbib,floatfix,superscriptaddress]{revtex4}

\usepackage{graphicx,bm,float, physics}
\usepackage{xcolor}

\makeatletter
\def\graphicscale{\twocolumn@sw{0.3}{0.4}}
\def\graphicthreescale{\twocolumn@sw{0.3}{0.4}}

\begin{document}

\title{Three-dimensional lattice SU($N_c$) gauge theories \\ 
with multiflavor scalar fields in the adjoint representation}

\author{Claudio Bonati} 
\affiliation{Dipartimento di Fisica dell'Universit\`a di Pisa 
       and INFN, Pisa, Italy}

\author{Alessio Franchi} 
\affiliation{Dipartimento di Fisica dell'Universit\`a di Pisa 
       and INFN, Pisa, Italy}
       
\author{Andrea Pelissetto}
\affiliation{Dipartimento di Fisica dell'Universit\`a di Roma Sapienza
        and INFN, Roma, Italy}

\author{Ettore Vicari} 
\affiliation{Dipartimento di Fisica dell'Universit\`a di Pisa
       and INFN, Pisa, Italy}

\date{\today}

\begin{abstract}
We consider three-dimensional lattice SU($N_c$) gauge theories with multiflavor
($N_f>1$) scalar fields in the adjoint representation.  We investigate their
phase diagram, identify the different Higgs phases with their gauge-symmetry
pattern, and determine the nature of the transition lines.  In particular, we
study the role played by the quartic scalar potential and by the gauge-group
representation in determining the Higgs phases and the global and gauge
symmetry-breaking patterns characterizing the different transitions. The
general arguments are confirmed by numerical analyses of Monte Carlo results
for two representative models that are expected to have qualitatively different
phase diagrams and Higgs phases.  We consider the model with $N_c = 3$, $N_f=2$
and with $N_c=2$, $N_f= 4$. This second case is interesting phenomenologically
to describe some features of cuprate superconductors.
\end{abstract}

\maketitle


\section{Introduction}
\label{intro}

Gauge symmetries represent a fundamental feature of high-energy
particle theories~\cite{Weinberg-book,Wilson-74,ZJ-book} and of
emerging phenomena in condensed matter
physics~\cite{Wegner-71,ZJ-book,Sachdev-19,Anderson-book}.  It is
therefore important to understand the role they play in gauge models.
In particular, it is crucial to have a solid understanding of how they
relate to global symmetries and of their role in determining the phase
structure of the model, the nature of its different Higgs phases and
of its quantum and thermal transitions.

We address these issues in three-dimensional (3D) lattice gauge models
with multicomponent scalar fields. We consider a lattice model with
O($N_f$) global invariance, SU($N_c$) local invariance, and in which
the scalar-matter field transforms in the adjoint representation of
SU($N_c$) and in the fundamental representation of O($N_f$)
~\cite{GG-72,FS-79}.  This model is of direct phenomenological
interest. In particular, the gauge model with $N_c=2$ and $N_f=4$
has been proposed as an effective model for optimal doping criticality
in hole-doped cuprate superconductors~\cite{SSST-19,SPSS-20}.

Studies addressing the interplay between global and gauge non-Abelian
symmetries in 3D models have been already reported.  We mention
Refs.~\cite{BPV-19,BPV-20} that studied models with fields
transforming under the fundamental representation of the gauge group:
Ref.~\cite{BPV-19} studied a model with a local SU($N_c$) and a global
SU($N_f$) invariance and Ref.~\cite{BPV-20} studied a model with
global O($N_f$) and local SO($N_c$) invariance.  Other studies have
focused on Abelian U(1) gauge theories~\cite{HLM-74,MZ-03}, such as
the lattice Abelian-Higgs model with
compact~\cite{PV-19-AH3d,WBJSS-05,KNS-02,MHS-02,KKLP-98} and
noncompact~\cite{BPV-21-nc,HBBS-13,KMPST-08,MV-08} gauge fields, and
with higher-charge scalar
fields~\cite{BPV-20-hc,WBJS-08,CIS-06,CFIS-05,NSSS-04,SSNHS-03,SSSNH-02}.

In this paper we extend these studies. First, we investigate the role
played by the gauge-group representation of the scalar fields. In
particular, we consider lattice SU($N_c$) gauge theories with
multiflavor scalar matter in the adjoint representation. Second, we
consider a generic quartic scalar potential, obtaining a richer phase
diagram with different Higgs phases.  We mention that some results for
this model have been already reported in Ref.~\onlinecite{SPSS-20},
which discusses the phase diagram and the different Higgs phases for
$N_c=2$ and $N_f =4$.  We extend here those results, presenting a
numerical analysis of the nature of the phase transitions along the
transition lines that separate the different phases.  We also mention
that the phase behavior of the same model has been studied also in two
dimensions~\cite{BFPV-21}, finding that the asymptotic
zero-temperature behavior (continuum limit) is the same as in models
defined on symmetric spaces that have the same global
symmetry~\cite{BHZ-80}.

The phase diagram of the lattice SU($N_c$) gauge model with multiflavor
scalar matter in the adjoint representation depends on the number of
colors $N_c$ and flavors $N_f$.  Its low-temperature Higgs phases
are essentially determined by the nature of the scalar configurations
in the low-temperature limit, and also by the topological properties
of the gauge fields. In particular, qualitatively different behaviors
emerge for $N_f\le N_c^2-1$ and $N_f>N_c^2-1$. In the first case there
is only one low-temperature Higgs phase, while the second case
presents various low-temperature Higgs phases.  Correspondingly, we
observe transitions that are related to the breaking of the global
symmetry group acting on the scalar fields, and topological
transitions separating phases with different topological properties of
the gauge field.  We present numerical studies based on Monte Carlo
simulations for one representative of each class of models. We study
the model for $N_c=3$ and $N_f=2$---in this case we have
$N_f<N_c^2-1$---and for $N_c=2$ and $N_f=4$, for which $N_f>N_c^2-1$.
Some numerical results for $N_c=2$ and $N_f=4$ in the strong
gauge-coupling limit were also reported in Ref.~\onlinecite{SPSS-20}.

The model with one scalar field, i.e., with $N_f=1$, is also
phenomenologically interesting---it is relevant for electron-doped
cuprates \cite{SSST-19}. However, its phase diagram is somewhat
trivial, as it presents a single thermodynamical phase, with two
continuously-connected regimes, a disordered-like and a Higgs-like
regime~\cite{DHKR-02,SSST-19}. Indeed, the existence of a distinct
low-temperature Higgs phase generally requires the breaking of a
global symmetry group, which is only possible for $N_f\ge 2$.

The paper is organized as follows. In Sec.~\ref{model} we define the
lattice SU($N_c$) gauge model with $N_f$ scalar fields in the adjoint
representation. In Sec.~\ref{obsfss} we introduce the observables and
discuss their finite-size scaling (FSS) behavior, which will be at the
basis of our numerical analyses. In Sec.~\ref{gausym} we determine the
minimum-potential configurations, which specify the different Higgs
phases, and characterize the global and gauge symmetry-breaking
patterns.  In Sec.~\ref{sft} we discuss the renormalization-group (RG)
flow of the statistical field theory that is associated with the
lattice model, focusing on the case $N_c=2$.  In Sec.~\ref{partlims}
we discuss some limiting cases, corresponding to simpler models for
which some features of the phase diagram are already known. The next
two sections are dedicated to the presentation of the numerical
results.  In Sec.~\ref{numresnf2nc3} we discuss the phase diagram of
the model with $N_c=3$ and $N_f=2$, which is a representative of
models with $N_f\le N_c^2-1$. Sec.~\ref{numresnf4nc2} reports a
numerical analysis of the more interesting case with $N_c=2$ and
$N_f=4$, for which $N_f>N_c^2-1$.  Finally, in Sec.~\ref{conclu} we
summarize and draw our conclusions.  Some details on the MC
simulations and numerical analyses are reported in App.~\ref{mcsim}.

\section{Lattice SU($N_c$) gauge models with adjoint scalar fields}
\label{model}

We consider lattice gauge models that are invariant under local
SU($N_c$) and global O($N_f$) transformations, with scalar fields that
transform under the adjoint representation of SU($N_c$) and under the
fundamental representation of the O($N_f$) group.  They are defined on
cubic lattices of linear size $L$ with periodic boundary conditions.
The fundamental variables are real matrices $\Phi^{af}_{\bm x}$, with
$a=1,...,N_c^2-1$ ({\em color} index) and $f=1,...,N_f$ ({\em flavor}
index), defined on the lattice sites, and gauge fields $U_{{\bm
    x},\mu} \in {\rm SU}(N_c)$ associated with the lattice
links~\cite{Wilson-74}.  The partition function is
\begin{eqnarray}
&&  Z = \sum_{\{\Phi,U\}} e^{-\beta H}\,,\qquad \beta=1/T\,,\label{partfunc}\\ 
&&  H  = H_K(\Phi,U) + H_V(\Phi) + H_G(U)\,,
\label{hgauge}
\end{eqnarray}
where the lattice Hamiltonian $H$ is the sum of the kinetic term $H_K$
of the scalar fields, of the local scalar potential $H_V$, and of the
pure-gauge Hamiltonian $H_G$.  As usual, we set the lattice spacing
equal to one, so that all lengths are measured in units of the lattice
spacing.

The kinetic term $H_K$ is given by
\begin{eqnarray}
  H_K(\Phi,U) =   - J
  {N_f\over 2} \sum_{{\bm x},\mu} {\rm Tr} \,\Phi_{\bm x}^t \,
  \widetilde{U}_{{\bm x},\mu} \, \Phi_{{\bm x}+\hat{\mu}}^{\phantom t}\,,
  \label{Kinterm}
\end{eqnarray}
where the matrix $\widetilde{U}^{ab}_{{\bm x},\mu}$ is the adjoint
representation of the original link variable $U_{{\bm x},\mu}$, 
explicitly defined as
\begin{equation}
  \widetilde{U}^{ab}_{{\bm x},\mu} = 2 \, {\rm Tr}(\,U^{\dagger}_{{\bm x},\mu}
  T^a U_{{\bm x},\mu} T^b\,)\,, 
\quad a,b=1,...,N_c^2-1\,,
\label{utilde}
\end{equation}
where $T^a$ are the $N_c^2-1$ generators in the fundamental
representation, normalized so that ${\rm Tr} \, T^a T^b =
\frac{1}{2}\delta^{ab}$. In the following we fix $J=1$, so that
energies are measured in units of $J$.

The scalar potential term $H_V$ is written as
\begin{eqnarray}
  &&H_V(\Phi) = \sum_{\bm x} V(\Phi_{\bm x})\,,\label{potential}\\
  &&V(\Phi)={r\over 2} \, {\rm Tr}\,\Phi^t\Phi + {u\over 4} \, \left( {\rm
    Tr}\,\Phi^t\Phi\right)^2 + {v\over 4} \, {\rm Tr}\,(\Phi^t\Phi)^2
  \,,\nonumber
\end{eqnarray}  
which is the most general quartic potential invariant under
O($N_f$)$\otimes$O($N_c^2-1$) transformations.  For $v=0$, the
symmetry group of $H_V(\Phi)$ is larger, namely, the O($M$) group with
$M=N_f(N_c^2-1)$.

Finally, the pure-gauge plaquette term reads
\begin{eqnarray}
&&H_G(U) =
  - {\gamma\over N_c} \sum_{{\bm x},\mu>\nu}  {\rm Re} \,
  {\rm Tr}\, \Pi_{{\bm x},\mu\nu}\,,
\label{plaquette}\\
&&\Pi_{{\bm x},\mu\nu}=
      U_{{\bm x},\mu} \,U_{{\bm x}+\hat{\mu},\nu} \,U_{{\bm
    x}+\hat{\nu},\mu}^\dagger \,U_{{\bm x},\nu}^\dagger \,,
\nonumber
\end{eqnarray}
where the parameter $\gamma$ plays the role of inverse gauge coupling.

The model is invariant under global O($N_f$) transformations,
$\Phi^{af} \to \sum_g O^{fg} \Phi^{ag}$, and under local SU($N_c$)
transformations
\begin{equation}
U_{{\bm x},\mu} \to V_{\bm x} U_{{\bm x},\mu} V_{{\bm x} + \hat\mu}^\dagger 
\qquad
\Phi^{af}_{\bm x} \to \sum_b \widetilde{V}_{\bm x}^{ab} \Phi^{bf}_{\bm x},
\end{equation}
where $V_{\bm x}$ is an SU($N_c$) matrix and $\widetilde{V}_{\bm x}$
is the corresponding matrix in the adjoint representation
[$\widetilde{V}$ can be obtained from $V$ using the analogue of
  Eq.~(\ref{utilde})].

In our study we focus on a representative model with 
fixed-length scalar fields $\Phi_{\bm x}$, satisfying 
\begin{equation}
{\rm Tr}\, \Phi_{\bm x}^t \Phi_{\bm x} = 2\,.
\label{trphi2}
\end{equation}
Formally, this model can be obtained by taking the limit
$u,r\to\infty$ keeping the ratio $r/u=-2$ fixed. The corresponding
lattice Hamiltonian reads
\begin{eqnarray}
&&H = - {N_f\over 2} \sum_{{\bm x},\mu} {\rm Tr} \,\Phi_{\bm x}^t \,
\widetilde{U}_{{\bm x},\mu} \, \Phi_{{\bm x}+\hat{\mu}}^{\phantom t}
\label{hfixedlength}  \\
&&\quad+ {v\over 4} \sum_{\bm x}  {\rm Tr}\,(\Phi_{\bm x}^t\Phi_{\bm x})^2
- {\gamma\over N_c} \sum_{{\bm x},\mu>\nu}  {\rm Re} \, {\rm Tr}\,
\Pi_{{\bm x},\mu\nu}\,.\nonumber
\end{eqnarray}
Models with generic values of $r$ and $u$ are expected to have the 
same qualitative behavior as this simplified model.

For $\gamma=0$ each matrix $U_{{\bm x},\mu}$ can be multiplied by an
arbitrary $({\bm x},\mu)$-dependent element of the gauge-group center
${\mathbb Z}_{N_c}$ without changing the Hamiltonian, thus implying
that the gauge group is SU($N_c$)$/{\mathbb Z}_{N_c}$. In particular,
this implies $\langle {\rm Tr}\,\Pi_{{\bm x},\mu\nu} \rangle=0$ for
$\gamma = 0$.  Note also that, for $N_c = 2$ and again for $\gamma =
0$, because of the isomorphism SU$(2)/{\mathbb Z}_{2}=$SO(3), we
recover an SO(3) gauge theory with scalar matter in the fundamental
representation.

For $\gamma\neq 0$, the gauge Hamiltonian breaks the previous
symmetry.  However, the Hamiltonian is still invariant under a
subgroup of those transformations. More precisely, it is invariant
under the transformations $U_{{\bm x},\mu}\to c(x_{\mu})U_{{\bm
    x},\mu}$, where $c(x_{\mu})$ is an element of the gauge-group
center that depends only on $x_{\mu}$ (the component $\mu$ of the
position vector). When this symmetry is not spontaneously broken,
Wilson loops obey the area law and color charges transforming in the
fundamental representation are confined.

Finally, for $\gamma\to\infty$, the link variables $U_{{\bm x},\mu}$
become equal to the identity, modulo gauge transformations.  Thus, one
recovers a matrix scalar model which is invariant under global
O($N_f$)$\otimes$O($N_c^2-1$) transformations [for $v=0$, the global
  symmetry group is O($M$) with $M=N_f(N_c^2-1)$].  This is strictly
true only for an infinite system. On a finite lattice with periodic
boundary conditions, it is not possible to set $U_{{\bm x},\mu}=1$ on
all links and therefore, one ends up with a scalar model with
SU($N_c$) (since the fields transform under the adjoint
representation, the group is more precisely SU($N_c$)/${\mathbb
  Z}_{N_c}$) fluctuating boundary conditions (see
Ref.~\cite{BPV-21-ccb} for a discussion in the context of U(1) gauge
models).

\section{Observables, order parameter and finite-size scaling}
\label{obsfss}

To investigate the phase diagram of the lattice SU($N_c$) gauge theory
(\ref{hfixedlength}), we consider the energy density and the specific
heat, defined as
\begin{eqnarray}\label{ecvdef}
E = -\frac{1}{3 L^3} \langle H \rangle\,,\quad C_V
=\frac{1}{L^3}\left( \langle H^2 \rangle - \langle H
\rangle^2\right)\,.
\end{eqnarray}

The critical properties of the scalar fields can be monitored by using
the correlation functions of the gauge-invariant bilinear operators
\begin{equation}
  B_{\bm x}^{fg} = {1\over 2} \sum_a \Phi_{\bm x}^{af} \Phi_{\bm x}^{ag}\,,
  \qquad
  Q_{\bm x}^{fg} = B_{\bm x}^{fg} - {1\over N_f} \delta^{fg}\,,
\label{qdef}
\end{equation}
which satisfy ${\rm Tr} \, B_{\bm x}=1$ and ${\rm Tr} \, Q_{\bm x}=0$
due to the fixed-length constraint.  The bilinear scalar operator
$Q_{\bm x}$ provides the natural order parameter for the breaking of
the global O($N_f$) symmetry.  As we use periodic boundary conditions
for all fields, translation invariance holds.  We define the
two-point correlation function
\begin{equation}
G({\bm x}-{\bm y}) = \langle {\rm Tr}\, Q_{\bm x} Q_{\bm y} \rangle\,,  
\label{gxyp}
\end{equation}
the corresponding susceptibility $\chi=\sum_{\bm x} G({\bm x})$, and
the second-moment correlation length
\begin{eqnarray}
\xi^2 = {1\over 4 \sin^2 (\pi/L)} {\widetilde{G}({\bm 0}) -
  \widetilde{G}({\bm p}_m)\over \widetilde{G}({\bm p}_m)}\,,
\label{xidefpb}
\end{eqnarray}
where $\widetilde{G}({\bm p})=\sum_{{\bm x}} e^{i{\bm p}\cdot {\bm x}}
G({\bm x})$ is the Fourier transform of $G({\bm x})$, and ${\bm p}_m =
(2\pi/L,0,0)$. We also consider RG-invariant quantities, such as the
Binder parameter
\begin{equation}
U = \frac{\langle \mu_2^2\rangle}{\langle \mu_2 \rangle^2} \,, \qquad
\mu_2 = \frac{1}{L^6}  
\sum_{{\bm x},{\bm y}} {\rm Tr}\,Q_{\bm x} Q_{\bm y}\,,
\label{binderdef}
\end{equation}
and 
\begin{equation}\label{rxidef}
R_{\xi}=\xi/L\,.
\end{equation}
At continuous transitions RG-invariant quantities, generically denoted
by $R$, scale as~\cite{PV-02}
\begin{eqnarray}
R(\beta,L) = f_R(X) +  L^{-\omega} g_R(X) + \ldots \,, \label{scalbeh}
\end{eqnarray}
where
\begin{equation}
  X = (\beta-\beta_c)L^{1/\nu}\,,
  \label{Xdef}
  \end{equation}
and next-to-leading scaling corrections have been neglected. The
function $f_R(X)$ is universal up to a multiplicative rescaling of its
argument, $\nu$ is the critical exponent associated with the diverging
correlation length, and $\omega$ is the exponent associated with the
leading irrelevant operator.  In particular, $U^*\equiv f_U(0)$ and
$R_\xi^*\equiv f_{R_\xi}(0)$ are universal, depending only on the
boundary conditions and aspect ratio of the lattice.  Since $R_\xi$
defined in Eq.~\eqref{rxidef} is an increasing function of $\beta$, we
can write
\begin{equation}\label{uvsrxi}
  U(\beta,L) = F(R_\xi) + O(L^{-\omega})\,,
\end{equation}
where $F(x)$ depends on the universality class, boundary conditions,
and lattice shape, without any nonuniversal multiplicative factor.
Eq.~\eqref{uvsrxi} is particularly convenient to test
universality-class predictions, as it permits a direct comparison of
results for different models without requiring a tuning of
nonuniversal parameters.

The Binder parameter $U$ is also useful to identify weak first-order
transitions, especially when large lattices are required to obtain
evidence of a finite latent heat or of a bimodal energy distribution.
Indeed, at a first-order transition, the maximum $U_{\rm max}$ of $U$
increases as the volume $L^3$, i.e.~\cite{CLB-86,VRSB-93,CPPV-04}
\begin{equation}\label{Ufirst}
U_{\rm max}= a\,L^3  + O(1)\,.
\end{equation}
This is the key point which distinguishes first-order from continuous
transitions. Indeed, at a continuous phase transition, $U$ is finite
as $L\to \infty$; at the critical point $U$ converges to a universal
value $U^*$, while the data of $U$ corresponding to different values
of $R_{\xi}$ collapse onto a scaling curve as the volume is increased.
Therefore, $U$ has a qualitatively different scaling behavior for
first- and second-order transitions.  The absence of a data collapse
in plots of $U$ versus $R_{\xi}$ may be considered as an early
indication of the first-order nature of the
transition~\cite{PV-19}. To identify the transition, one can also
consider the specific heat. At first-order transitions, its maximum
value $C_{\rm max}(L)$ asymptotically increases as \cite{CLB-86}
\begin{eqnarray}
C_{\rm max}(L) = {\Delta_h^2\over 4}\,L^3 + O(1)\,,
  \label{cmaxsc}
\end{eqnarray}  
where $\Delta_h$ is the latent heat, defined as the difference $\Delta_h =
E(\beta\to\beta_c^+) - E(\beta\to\beta_c^-)$. Moreover, the value of
$\beta$ corresponding to the maximum converges to the critical value
$\beta_c$ as $\beta_{{\rm max},C}(L)-\beta_c\approx c\,L^{-3}$. Note
that $C_{\rm max}(L)$ may also diverge at continuous transitions (this
occurs when $\alpha > 0$), and therefore the identification of the
order of the transition from the behavior of $C_{\rm max}(L)$ requires
a detailed analysis of its asymptotic large-$L$ behavior.

\section{Higgs phases}
\label{gausym}

The lattice gauge models we consider may have different Higgs phases
associated with different symmetry-breaking patterns. They are
determined by the minima of the local scalar potential
(\ref{potential}), namely
\begin{eqnarray}
V(\Phi) = {v\over 4} \,{\rm Tr}\,(\Phi^t\Phi)^2\,,
\label{locpotential}
\end{eqnarray}
in the fixed-length limit ${\rm Tr}\, \Phi^t \Phi = 2$.  In the
following we summarize (using the notations of Ref.~\cite{BFPV-21})
the main properties of these phases, which crucially depend on the
number of colors $N_c$, of flavors $N_f$, and on the parameter
$v$~\cite{SPSS-20,SSST-19,BFPV-21}.  Moreover, as we shall see, their
nature may also depend on the behavior of the fluctuations of
variables associated with the gauge-group center ${\mathbb Z}_{N_c}$,
which are expected to undergo a transition at finite values of
$\gamma>0$.

\subsection{The model for $v<0$}
\label{negvcase}

For $v<0$ the mininum-potential configurations can be generally written
as~\cite{SSST-19,BFPV-21}
\begin{equation}
  \Phi^{af} = \sqrt{2} \, s^a z^f\,, 
\label{Fieldbetainf-1}
\end{equation}
where ${\bm s}$ and ${\bm z}$ are unit real vectors of dimension
$N^2_c - 1$ and $N_f$, respectively. To identify the symmetry breaking
pattern at the transition, we should identify the stabilizer group
(little group in Wigner's notation) of the solution
(\ref{Fieldbetainf-1}), i.e., the group of O($N_f$) transformations
that leave the field (\ref{Fieldbetainf-1}) invariant, modulo gauge
transformations. Explicitly, we should find the orthogonal matrices
$O^{fg}$ such that
\begin{equation}
  \sum_{g} O^{fg} s^a z^g = \sum_b \widetilde{V}^{ab} s^b z^f\,,
  \label{stabcond}
\end{equation}
for some $V\in\hbox{SU}(N_c)$ (the tilde accent indicates the adjoint
representation). It is immediate to verify that $V$ should satisfy
$|\sum_{ab} s^a \widetilde{V}^{ab} s^b| = 1$, so that
Eq.~(\ref{stabcond}) can be written as
\begin{equation}
\sum_{g} O^{fg} z^g = \pm z^f\,. 
\end{equation}
The invariance group is therefore ${\mathbb Z}_2 \otimes O(N_f-1)$ and
the global symmetry breaking pattern is
\begin{equation}
O(N_f) \to {\mathbb Z}_2 \otimes O(N_f-1)\,.
\label{sbp-vlt0}
\end{equation}
We also define a gauge-symmetry breaking pattern as the stabilizer 
of the minimum-potential solution with respect to the gauge group.
For this purpose, we determine the matrices 
$V\in \hbox{SU}(N_c)$ such that 
\begin{equation}
\sum_b \widetilde{V}^{ab} s^b = s^a\,.
\end{equation}
Defining $\hat{T} = \sum_a s^a T^a$, and using Eq.~(\ref{utilde})
we obtain 
\begin{equation}
2 \sum_{a} T^a \hbox{Tr}\, (V^\dagger T^a V \hat{T}) = \hat{T}\,.
\end{equation}
Using the completeness relation for the generators, we end up with the
condition $[V,\hat{T}] = 0$.  The stabilizer subgroup is therefore
${\rm U}(1)\oplus {\rm U}(N_c-2)$, so that for $v<0$ we observe a
gauge symmetry breaking pattern
\begin{equation}
{\rm SU}(N_c) \to {\rm U}(1)\otimes{\rm U}(N_c-2)\,,
\label{gspnegv}
\end{equation}
independently of the flavor number $N_f$. In particular, for $N_c=2$,
we have~\cite{SPSS-20} SU(2)$\to$U(1) [equivalently, disregarding
  discrete subgroups, it corresponds to O(3)$\to$O(2)].  Note that we
are not claiming here that the gauge symmetry is broken in the
standard statistical-mechanics sense (i.e., that we can force the
system in one specific minimum, for instance, by appropriately fixing
the boundary conditions), as this is forbidden by well-known rigorous
arguments \cite{Elitzur-75,DFG-78,BN-87}.  The right-hand side of the
gauge-symmetry breaking pattern only represents the residual gauge
symmetry of the minimum-potential configuration once the scalar fields
have been fixed to a specific value by means of an appropriate
gauge-fixing condition (see Ref.~\cite{BN-87} for a discussion of the
role of gauge fixings), i.e., once a specific value of $s$ in
Eq.~(\ref{Fieldbetainf-1}) has been chosen.

One may also establish a correspondence between the critical behavior
of the SU($N_c$) gauge model (\ref{hfixedlength}) and the 3D
RP$^{N_f-1}$ model~\cite{BFPV-21}.  Consider indeed the limit $v\to
-\infty$ at fixed $\beta$ and $J$.  In this limit $B_{\bm x}$, defined
in Eq.~(\ref{qdef}), becomes
\begin{equation}
B_{\bm x}^{fg} = z_{\bm x}^f z_{\bm x}^g\,, 
\label{bproj}
\end{equation}
i.e., it corresponds to a local projector onto a one-dimensional
subspace.  If we now assume that the dynamics in the gauge model is
determined by the fluctuations of the order parameter $B_{\bm x}$, or
equivalently of $Q_{\bm x}$, we identify the effective scalar model as
the RP$^{N_f-1}$ model. Indeed, the standard nearest-neighbor
RP$^{N-1}$ action is obtained by taking the simplest action for a
local projector $P^{fg}_{\bm x}$:
\begin{eqnarray}
  H_{\rm RP} = - J \sum_{{\bm x},\mu} \hbox{Tr}\, P_{\bm x} P_{{\bm
      x}+\hat{\mu}}\,, \qquad P_{\bm x}^{fg} = \varphi_{\bm x}^f
  \varphi_{\bm x}^g\,,
\label{srp}
\end{eqnarray}
where $\varphi^a_{\bm x}$ is a unit vector. We do not expect the limit
$v\to -\infty$ to be relevant. The crucial property should be the
structure of the low-temperature configurations, and thus we expect
RP$^{N_f-1}$ in the whole phase in which the symmetry-breaking
patterns (\ref{sbp-vlt0}) and (\ref{gspnegv}) hold.  We recall that 3D
RP$^{N-1}$ models undergo continuous transitions only for $N=2$---they
belong to the XY universality class. For any $N>2$, transitions are of
first order, as predicted by the Landau-Ginzburg-Wilson (LGW)
theory~\cite{PTV-18,FMSTV-05}.

\subsection{The model for $v>0$}
\label{posvcase}

The behavior of the model is more complex for $v>0$. The 
minimum-potential configurations
can be  parametrized as~\cite{SSST-19,BFPV-21}
\begin{equation}
\Phi^{ag} = 
\sqrt{2\over q} \sum_{k=1}^q C^{ak} F^{kg}\,, \quad
  q={\rm Min}[N_f,N_c^2-1]\,,
\label{Phi-vgt0}
\end{equation}
where $C$ and $F$ are orthogonal matrices of dimension $N_c^2-1$ and
$N_f$, respectively. To further simplify this expression we should
distinguish two cases: $N_f\le N_c^2-1$ and $N_f>N_c^2-1$

For $N_f\le N_c^2-1$, we can simplify Eq.~(\ref{Phi-vgt0}) into
\begin{equation}
  \Phi^{ag} = \sqrt{2\over q}\,C^{ag}\,,\qquad C\in {\rm O}(N_c^2-1)\,.
  \label{nfltncstuff}
  \end{equation}
Moreover, for $N_c=2$, since the adjoint representation of SU(2) is
equivalent to SO(3), one may further simplify the representation of
the mininum-potential configuration: all such configuration are
obtained by applying gauge transformations to
$\Phi^{ag}=\sqrt{2/q}\,\delta^{ag}$.

The global invariance group of
the ordered phase is given by those transformations
$O\in\hbox{O}(N_f)$ such that
\begin{equation}
\sum_g O^{fg} C^{ag} = \sum_b \widetilde{V}^{ab} C^{bf}, 
\end{equation}
for some SU($N_c$) matrix $V$.  This condition implies that $O^{fg} =
(C^t \widetilde{V}^t C)^{fg}$. Since the matrix $C^t \widetilde{V}^t
C$ is an element of the adjoint representation of SU($N_c$), $O$
should be an $N_f\times N_f$ submatrix of an element of
$\hbox{SU}(N_c)_{\rm adj}$. We write the corresponding global symmetry
breaking pattern as
\begin{equation}
\hbox{O}(N_f) \to \hbox{O}(N_f) \cap \hbox{SU}(N_c)_{\rm adj} .
\end{equation}
For $N_f = 2$ and 3, since $\hbox{SU}(N_c)_{\rm adj}$ includes
$\hbox{SU}(2)_{\rm adj} = \hbox{SO}(3)$ and inversion transformations
on the first $N_f$ components, we have $\hbox{O}(N_f) \cap
\hbox{SU}(N_c)_{\rm adj} = \hbox{O}(N_f)$.  Thus, there is no global
symmetry breaking, and thus no transition is expected.  These
conclusions are consistent with the more general argument presented in
Ref.~\cite{BFPV-21}.  It was noted that the gauge-invariant order
parameter $Q$ defined in Eq.~(\ref{qdef}) vanishes if the fields are
given by Eq.~(\ref{Phi-vgt0}), as it does in the disordered phase.
Thus, the system is not expected to have low-temperature phases in
which the gauge-invariant bilinear operator $Q$ condenses.

For $N_f>N_c^2-1$, the minimum-potential configurations can be
parametrized as \cite{BFPV-21}
\begin{equation}
  \Phi^{ag} = \sqrt{2\over q}\,F^{ag}\,,\qquad F\in {\rm O}(N_f)\,.
  \label{nfgtncstuff}
\end{equation}
Modulo global O($N_f$) transformations, a simple representative is
$\Phi^{ag} = \sqrt{2/q}\,\delta^{ag}$. For what concerns the global-symmetry
breaking pattern, the transformations $O\in\hbox{O}(N_f)$ that leave
the minimum-potential configurations invariant modulo gauge
transformations satisfy the condition $ O^{ab} = \widetilde{V}^{ab}
$, so that the symmetry-breaking pattern is
\begin{equation}
\hbox{O}(N_f) \to \hbox{O}(N_f - N^2_c + 1) \oplus \hbox{SU}(N_c)_{\rm adj}\,.
\label{gsbpvp1}
\end{equation}
For $N_c =2$, it becomes 
\begin{equation}
\hbox{O}(N_f) \to \hbox{O}(N_f - 3) \oplus \hbox{SO}(3)\,.
\label{gsbpvp2}
\end{equation}
If we additionally set $N_f = 4$, it becomes $\hbox{O}(4) \to
\hbox{O}(3)$, which is the symmetry breaking pattern of the O(4)
vector model.  If we consider the gauge group, instead, since 
the only matrix that leaves $\Phi^{ag} =
\sqrt{2/q}\,\delta^{ag}$ invariant
is $\widetilde{V}=1$, the stabilizer group is the center
${\mathbb Z}_{N_c}$. The gauge symmetry breaking pattern is therefore
\begin{equation}
\hbox{SU}(N_c) \to {\mathbb Z}_{N_c}\,.
\label{gausympat}
\end{equation}
In the previous discussion we have characterized the phases on the
basis of the different minima of the potentials. However, phases may
also depend on topological properties of the gauge fields, which are
controlled by the coupling $\gamma$. In particular, the modes related
to the center of the gauge group ${\mathbb Z}_{N_c}$ may undergo a
confining-deconfining phase transition at finite values of $\gamma$,
giving rise to low-temperature Higgs phases that have the same global
and local gauge symmetry breaking patterns, but differing for the
topological nature of the gauge-center excitations.  We expect these
phenomena to be relevant for $v > 0$, when the gauge symmetry breaking
pattern is ${\rm SU}(N_c) \rightarrow {\mathbb Z}_{N_c}$, so that the
minimum-potential configurations are only invariant under the
gauge-group center.

To understand the role of the gauge-group center, we consider the
limit $\beta\to\infty$ keeping $\kappa\equiv \beta\gamma$ fixed. In
this limit, the relevant configurations minimize the potential and the
scalar kinetic energy $H_K$.  As discussed in Ref.~\cite{BFPV-21}, for
$v>0$ the minimization of $H_K$ implies $\widetilde{U}_{{\bm
    x},\mu}=1$, so that $U_{{\bm x},\mu} = \lambda_{{\bm x},\mu}\in
{\mathbb Z}_{N_c}$.  In this limit, model (\ref{hfixedlength}) reduces
to the lattice ${\mathbb Z}_{N_c}$ gauge theory
\begin{eqnarray}
H_{{\mathbb Z}_{N_c}} = - \kappa \sum_{{\bm x},\mu>\nu} {\rm Re} \,
  \lambda_{{\bm x},\mu} \,\lambda_{{\bm x}+\hat{\mu},\nu} \,
  \bar\lambda_{{\bm x}+\hat{\nu},\mu} \,\bar\lambda_{{\bm x},\nu} \, .
  \label{zncgauge}
\end{eqnarray}
In three dimensions, this lattice discrete gauge model undergoes a
continuous transition at a finite $\kappa_c$ (see
Sec~\ref{betainfkappa} for more details). For example, for $N_c = 2$
the Hamiltonian (\ref{zncgauge}) corresponds to a lattice ${\mathbb
  Z}_{2}$ gauge theory~\cite{Wegner-71}, which presents a
small-$\kappa$ confined phase and a large-$\kappa$ deconfined phase
(which may carry topological order at the quantum
level~\cite{Sachdev-19}), separated by a critical point at $\kappa_c =
0.761413292(11)$ (see Sec.~\ref{betainfkappa}).  If the ${\mathbb
  Z}_{N_c}$ gauge transition persists for finite values of $\beta$,
then, when varying $\gamma$, we may have different low-temperature
Higgs phases that are associated with the same gauge-symmetry pattern
${\rm SU}(N_c) \rightarrow {\mathbb Z}_{N_c}$ but that differ in the
large-scale behavior of the ${\mathbb Z}_{N_c}$ variables.  This may
lead to a change of the nature of the phase transition from the
disordered to the Higgs phases when $N_f>N_c^2-1$, or give rise to
observable effects on the scalar correlations for $N_f\le N_c^2-1$,
which are not expected to order for $v>0$.

\section{RG flow of the gauge field theory}
\label{sft}

In this section we discuss the RG flow of the statistical field theory
corresponding to the lattice model (\ref{hgauge}), focusing on the
case $N_c=2$. The starting point is a scalar theory in which the
fundamental field is a real matrix $\Phi_{af}$ ($a=1,...,N_c^2-1$ and
$f=1,...,N_f$), transforming as the corresponding lattice field (see
Sec.~\ref{model}). The corresponding Hamiltonian includes all field
monomials of dimension less or equal to four that are invariant under
global O($N_f$)$\otimes$O($N_c^2-1$) transformations.  To obtain a
model invariant under SU($N_c$) gauge transformations, we add an
SU($N_c$) gauge field $A_{\mu}^a$ and set ${\cal A}_{\mu\,ab} \equiv i
A_{\mu}^k T_{A,ab}^{k}$, where $T_{A,ab}^{k}=-i f^{abk}$ are the
SU($N_c$) generators in the adjoint representation ($f^{abc}$ are the
structure constants of the SU($N_c$) group). The Hamiltonian density
is
\begin{eqnarray}
&&{\cal H}= {1\over 4 g_0^2}(F^k_{\mu\nu})^2 + 
(\partial_\mu \Phi_{af} + {\cal A}_{\mu\,ab} \Phi_{bf})^2 
  + {1\over 2} r \,{\rm Tr}\,\Phi^t\Phi
\nonumber  \\
&&\;\;+ {1\over 4} u_0 ({\rm Tr}\,\Phi^t\Phi)^2 + {1\over 4} v_0 \left[ 
{\rm Tr}\,(\Phi^t\Phi)^2   - ({\rm Tr}\,\Phi^t\Phi)^2
\right]\qquad
\label{cogau}
\end{eqnarray}
where $F_{\mu\nu}^k$ is the non-Abelian field strength associated with
the gauge field $A_{\mu}^k$. To determine the nature of the
transitions described by the continuum SU($N_c$) gauge theory
(\ref{cogau}), one studies the RG flow determined by the $\beta$
functions of the model in the coupling space.

In the $\epsilon$-expansion framework, the RG flow
close to four dimensions is determined by the 
one-loop $\overline{\rm MS}$ $\beta$ functions.
Introducing the
renormalized couplings $u$, $v$, and $\alpha = g^2$, the one-loop
$\beta$ functions for $N_c=2$ are given by~\cite{SSST-19}
\begin{eqnarray}
\beta_u &=& -\epsilon u + {3 N_f + 8\over 6} u^2   
\label{betas}\\
&&+ {N_f-1\over 3} (v^2 - 2 u v)
- 3 u \alpha  + {9\over 4}\alpha^2 \,,
\nonumber\\
\beta_v &=& - \epsilon v +
{N_f-5\over 6}v^2 + 2uv - 3 v\alpha + {9\over 8} \alpha^2\,,
\nonumber\\
\beta_\alpha &=&
-\epsilon \alpha  + {N_f  - 22 \over 12}\alpha^2
\,,\nonumber
\end{eqnarray}
where $\epsilon\equiv 4-d$. The normalizations of the couplings can be
easily inferred from the above expressions.\footnote{The
$\beta$-functions (\ref{betas}) must be equal to those of the SO(3)
gauge theory in the fundamental representation. They indeed agree for
$N_c=2$ with those of the SO($N_c$) gauge model reported below:
\begin{eqnarray}
\beta_u &=& -\epsilon u + {N_c N_f + 8\over 6} u^2   
+ {(N_f-1)(N_c-1)\over 6} (v^2 - 2 u v) \nonumber\\
&&\;- {3\over 2} (N_c-1)u \alpha  + {9\over 8}(N_c-1)\alpha^2 \,,
\nonumber\\
\beta_v &=& - \epsilon v +
     {N_c+N_f-8\over 6}v^2 + 2uv
     \nonumber\\
&&\; - {3\over 2}(N_c-1) v\alpha 
    + {9\over 8} (N_c-2) \alpha^2\,,
\nonumber\\
\beta_\alpha &=&
-\epsilon \alpha  + {N_f  - 22(N_c-2) \over 12}\alpha^2
\,.\nonumber
\end{eqnarray}
We report them here, as a few misprints are present in the expressions
reported in Ref.~\onlinecite{BPV-20}. } The $\beta$-functions
(\ref{betas}) have a stable fixed point for sufficiently large $N_f$,
more precisely for $N_f > N^* + O(\epsilon)$ with $N^*\approx 210.5$.
In particular, in the large-$N_f$ limit the $\beta$ functions can be
written in terms of the large-$N_f$ parameters $\hat{u}\equiv N_f u$,
$\hat{v}\equiv N_f v$, $\hat{\alpha}\equiv N_f \alpha$, as
\begin{eqnarray}
&&  \beta_{\hat{u}} = -\epsilon \hat{u} + {1\over 6} \hat{u}^2 +
       {1\over 3}(\hat{u}-\hat{v})^2 \,,  \label{largenfbeta}\\
&&\beta_{\hat{v}} = - \epsilon \hat{v} +
{1\over 6}\hat{v}^2 \,, \quad
\beta_{\hat{\alpha}} =
-\epsilon \hat{\alpha}  + {1\over 12}\hat\alpha^2
\,,\nonumber
\end{eqnarray}
which have a stable fixed point located at
\begin{eqnarray}
  \hat\alpha^* = 12\epsilon\,,\quad
  \hat{u}^* = 6\epsilon\,,\quad
    \hat{v}^* = 6\epsilon\,.
  \label{fpln}
\end{eqnarray}
Note that the stable fixed point in the large-$N_f$ limit is located
in the region $v>0$. Thus, it should describe the continuous
transitions between the disordered phase and the positive-$v$ Higgs
phase discussed in Sec.~\ref{posvcase}.

\section{Some particular cases}
\label{partlims}

In this section we discuss some particular cases of the gauge model
(\ref{hfixedlength}), which correspond to lattice models that have
already been studied in the literature. This analysis will provide us
some indications on the phase diagram of the full theory.

\subsection{The model for $N_c=2$, $\gamma=0$, and $v=0$}
\label{nc2ga0}

For $N_c=2$ and $\gamma=0$, the model (\ref{hfixedlength}) is
equivalent to a lattice SO(3) gauge model with $N_f$ scalar flavors in
the fundamental representation. The Hamiltonian is
\begin{equation}
\label{hfixedlengthso3}
H=- {N_f\over 2} \sum_{{\bm x},\mu} {\rm Tr} \,\Phi_{\bm x}^t \,
V_{{\bm x},\mu} \, \Phi_{{\bm x}+\hat{\mu}}^{\phantom t}
+ {v\over 4} \sum_{\bm x}  {\rm Tr}\,(\Phi_{\bm x}^t\Phi_{\bm x})^2\,,
\end{equation}
where the link variables $V_{{\bm x},\mu}$ belong to the fundamental
representation of the gauge group SO(3). For $v=0$ this model was
discussed in Ref.~\cite{BPV-20}. It was predicted that, for any $N_f$,
the system undergoes a finite-temperature transition which is the same
as in the corresponding RP$^{N_f-1}$ model,
cf. Eq.~(\ref{srp}). Therefore, one predicts a continuous XY
transition for $N_f=2$ and a first-order transition for any $N_f > 2$.
Indeed, in the LGW Hamiltonian appropriate for the RP$^{N_f-1}$
model~\cite{PTV-18,FMSTV-05}, a cubic $\Phi^3$ term is always present
for $N_f > 2$, a presence which is usually considered as an indication
of a first-order transition for 3D statistical models.

These predictions have been confirmed numerically \cite{BPV-20}.  For
$N_f=2$ there is a continuous XY transition at $\beta_c=1.97690(7)$,
while for $N_f=3$ there is a first-order transition at $\beta_c
\approx 1.77$.  These numerical results indicate that, for $v = 0$,
the relevant low-temperature configurations are those of the form
(\ref{Fieldbetainf-1}), that correspond to the minima of the potential
for $v < 0$.

\subsection{The limit $\gamma\to\infty$}
\label{gammainfty}

For $\gamma\to\infty$ the variables $U_{{\bm x},\mu}$ converge 
to the identity, apart from gauge transformations. Thus, we obtain the 
scalar model
\begin{eqnarray}
  H = - {N_f\over 2} \sum_{{\bm x},\mu}
  {\rm Tr} \,\Phi_{\bm x}^t
  \, \Phi_{{\bm x}+\hat{\mu}}^{\phantom t}
  + {v\over 4} \sum_{\bm x}  {\rm Tr}\,(\Phi_{\bm x}^t\Phi_{\bm x})^2
  \,,\quad
\label{hfixedlengthgammainf}  
\end{eqnarray}
with a global O($N_f$)$\otimes$O($N_c^2-1$) symmetry.  For $v = 0$,
the symmetry group is larger, namely O($M$) with $M=N_f(N_c^2-1)$, and
therefore we expect continuous transitions belonging to the O($M$)
vector universality class. For $v\neq 0$, the models
(\ref{hfixedlengthgammainf}) may undergo a finite-temperature
continuous transition only if a corresponding universality class
exists and, in particular, only if the corresponding LGW $\Phi^4$
theory has a stable fixed point.  RG analyses indicate that continuous
transitions are possible for $N_f=2$ and $N_c=2$
~\cite{Kawamura-98,PRV-01,PV-02,Parruccini-03,CPPV-04,
  DPV-04,NO-14,HKS-20}, for both $v<0$ and $v>0$, and
for~\cite{DPV-06} $N_f=4$ and $N_c=2$ when $v<0$. Moreover, for $v>0$
there is a stable fixed point for sufficiently large $N_f$ at fixed
$N_c$ and sufficiently large $N_c$ at fixed $N_f$ (in particular for
$N_c=2$ and any $N_f$)~\cite{PRV-01-ln,Kawamura-98,CPPV-04}. It is not
clear whether the fixed points of the O($N_f$)$\otimes$O($N_c^2-1$)
field theory are relevant for the behavior for finite values of
$\gamma$. For instance, the O($M$) fixed point that controls the
behavior for $\gamma=\infty$ and $v=0$ is unstable with respect to the
gauge coupling, and is therefore irrelevant for the finite-$\gamma$
behavior, although it is expected to give crossover effects for large
values of $\gamma$. There are at present no analogous results for
$v\not=0$.

\subsection{The limit $\beta\to\infty$}
\label{betainf}

In the limit $\beta\to\infty$ the behavior of the system is controlled
by the configurations minimizing the Hamiltonian.  As already
discussed in Sec.~\ref{gausym}, two different low-temperature phases
occur for $v<0$ and $v>0$.  Therefore, in this limit we expect a
transition line for $v=0$ and any $\gamma$.  The transition line
should be of first order for any $N_f$ and $N_c$, as it separates
phases that correspond to different minima of the potential.

\subsection{The limit $\beta\to\infty$ keeping
  $\kappa\equiv\beta\gamma$ fixed}
\label{betainfkappa}

Let us now consider the limit $\beta\to\infty$ keeping $\kappa\equiv
\beta\gamma$ fixed. As mentioned in Sec.~\ref{posvcase}, for $v>0$ the
model (\ref{hfixedlength}) reduces to the lattice ${\mathbb Z}_{N_c}$
gauge theory defined in Eq.~(\ref{zncgauge}).  This model can be
related by duality to the $N_c$-state clock spin
model~\cite{SSNHS-03}, characterized by a global ${\mathbb Z}_{N_c}$
symmetry. For $q=2$, the $q$-state clock model is equivalent to the
standard Ising model and thus we expect an Ising transition.  Duality
allows us to obtain $\kappa_c$ for $N_c=2$: $\kappa_c = {1\over 2} \ln
\coth \beta_{I,c}$, where $\beta_{I,c}$ is the inverse temperature of
the Ising model. Using~\cite{FXL-18} $\beta_{I,c} = 0.221654626(5)$,
we obtain $\kappa_c = 0.761413292(11)$. For $q=3$, the $q$-state clock
model is equivalent to a three-state Potts model, which can only
undergo first-order transitions.  For larger values of $q$, we expect
a continuous transition.  It belongs to the Ising universality class
for~\cite{HS-03} $q=4$, and to the 3D XY universality class
for~\cite{HS-03,Hasenbusch-19,PSS-20} $q\ge 5$.  Note, however, that
in the $q\to\infty$ limit we recover the pure U(1) gauge theory, with
$\lambda_{{\bm x},\mu}\in {\rm U}(1)$, which is known~\cite{QED} to have no
transitions for finite values of $\kappa$. Therefore, if a transition
occurs for any finite $q$, we must have $\kappa_c\to\infty$ in the
$q\to\infty$ limit.

Since for $v>0$ and $N_f>N_c^2-1$ the low-temperature Higgs phase is
characterized by the gauge-symmetry breaking pattern ${\rm SU}(N_c)
\rightarrow {\mathbb Z}_{N_c}$ (see Sec.~\ref{posvcase}), it seems
natural to expect that the confinement-deconfinement center transition
also persists for finite $\beta$, giving rise to two different
positive-$v$ Higgs phases, depending on $\gamma$.

For $v<0$, the low-temperature Higgs phase is characterized by a residual
continuous gauge symmetry, see Eq.~(\ref{gspnegv}).  Since 3D pure gauge
theories with continuous gauge group do not display any
confinement-deconfinement transition, the same is expected to happen for the
model (\ref{hfixedlength}) when $v<0$.

\subsection{The limit $v\to\pm\infty$}
\label{modvinf}

For $|v|\to \infty$, configurations are constrained to be minima of
the the scalar potential (\ref{locpotential}). For $v \to +\infty$,
the scalar fields take the form (\ref{Phi-vgt0}), reducing the model
to a particular $\sigma$ model. Transitions are expected for
$N_f>N_c^2-1$, with the global symmetry-breaking pattern
(\ref{gsbpvp1}) [or (\ref{gsbpvp2}) for $N_c =2$].  For $N_f=4$,
$N_c=2$, the global symmetry-breaking pattern is O(4)$\to$O(3) and
therefore the transition should belong to the O(4) vector universality
class.

For $v\to -\infty$ scalar variables take the form (\ref{bproj}). As
discussed in Sec.~\ref{negvcase}, one expects to recover the effective
RP$^{N_f-1}$ model (\ref{srp}), whose transitions are continuous for
$N_f=2$ and of first order for any $N_f>2$.

\section{Results for $N_c=3$ and $N_f=2$}
\label{numresnf2nc3}

\begin{figure}[tbp]
\includegraphics[width=0.95\columnwidth, clip]{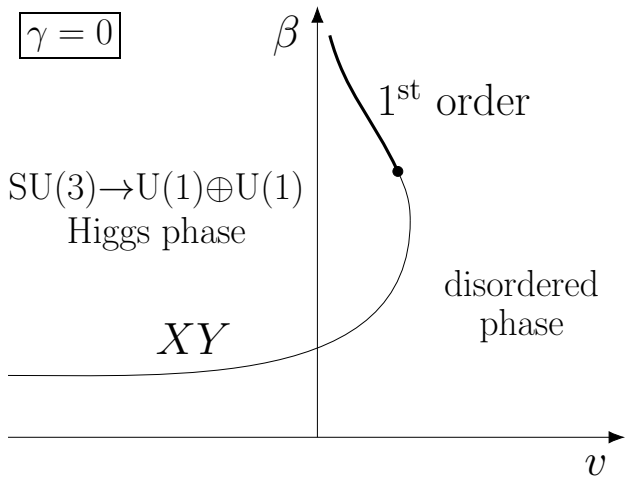}
\caption{A sketch of the phase diagram for $N_c=3$, $N_f=2$, and
  $\gamma =0$, inferred from the numerical results.  }
\label{phdianc3nf2}
\end{figure}

In this section we determine the phase diagram for $N_f=2$, $N_c=3$,
and $\gamma=0$.  According to the discussion reported in
Sec.~\ref{betainf}, since $N_f < N^2_c - 1$, for $\beta = \infty$
there is only one ordered Higgs phase, which is obtained for $v < 0$.
For finite values of $\beta$ we expect therefore only two phases: a
disordered phase and an ordered Higgs phase, separated by a single
transition line.  As discussed in Sec.~\ref{gausym}, the transitions
between the disordered and Higgs phases should be described by an
effective RP$^1$ model, which is equivalent to the XY model for
${\mathbb Z}_2$ gauge-invariant observables. Therefore, such
transitions should belong to the XY universality class~\cite{PV-02},
if they are continuous.  The transition line is expected to approach
the point $v=0$ in the $\beta=+\infty$ limit.  Moreover, since this
ending point should correspond to a first-order transition as outlined
in Sec.~\ref{betainf}, we expect the transition line to become of
first order for large values of $\beta$. The phase diagram obtained
from our MC simulations, see Fig.~\ref{phdianc3nf2}, is fully
consistent with these considerations.  Note that the transition line
intersects the line $v = 0$ at a finite $\beta$ value, so that the
ordered Higgs phase is also present for finite-$\beta$ positive values
of $v$. Of course, the transition line should be reentrant, since
$v_c\to 0^+$ for $\beta\to\infty$.

To verify the phase diagram sketched in Fig.~\ref{phdianc3nf2}, we
have performed simulations for $v = 0$ varying $\beta$, and at fixed
$\beta$ (we have considered $\beta = 5.2$, 6, 7.5, 9, 12) varying
$v$. We have verified the reentrant nature of the transition line and
that the transition changes from a continuous one to a first-order one
as $\beta$ increases (the tricritical point, where the order of the
transition changes, should satisfy $6 \lesssim \beta_{\rm tri}
\lesssim 7.5$).  Some technical details on the MC simulations are
reported in App.~\ref{mcsim}.

\begin{figure}[tbp]
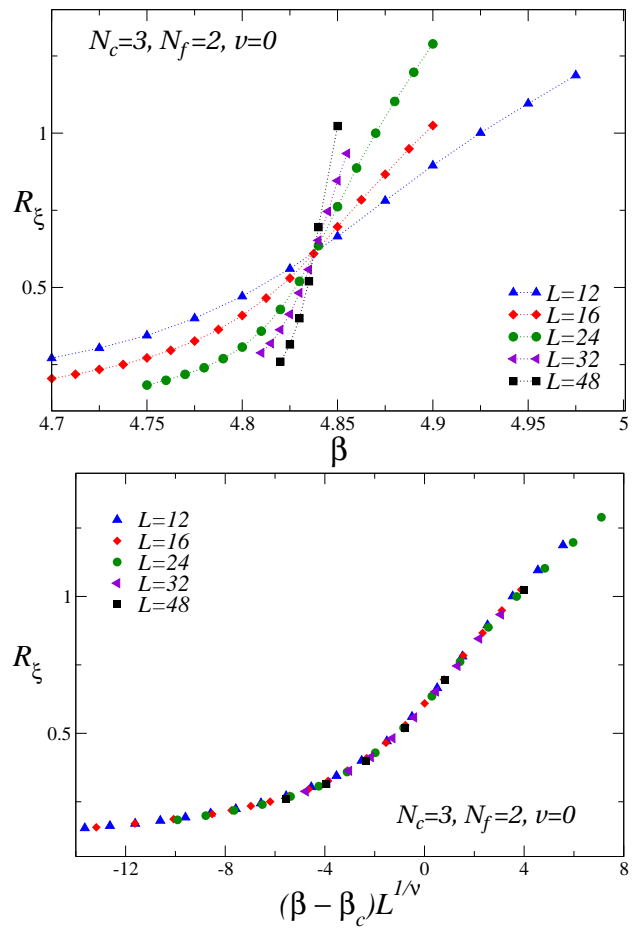

\includegraphics[width=0.95\columnwidth, clip]{sun3c2f_rxi_beta.eps}
\includegraphics[width=0.95\columnwidth, clip]{rxi_scaling.eps}
\caption{Top: Plot of $R_\xi$ versus $\beta$ for $N_c=3$, $N_f=2$,
  $\gamma=0$, and $v=0$. Bottom: Plot of $R_\xi$ versus $(\beta -
  \beta_c)L^{1/\nu}$, using the XY correlation-length exponent
  $\nu=0.6717$ and $\beta_c=4.8374$. Data collapse on an asymptotic
  curve with increasing $L$. }
\label{3col2flav_v0}
\end{figure}

\begin{figure}[tbp]
\includegraphics[width=0.95\columnwidth, clip]{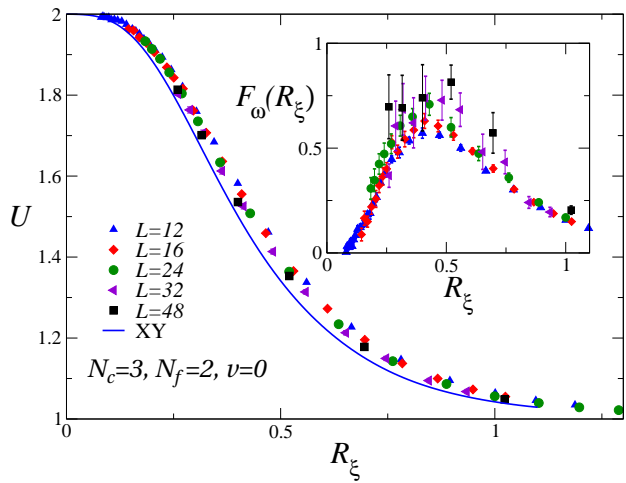}
\caption{Plot of $U$ versus $R_\xi$ for $N_c=3$, $N_f=2$, $\gamma=0$,
  and $v=0$. The continuous line represents the universal curve
  $F(R_\xi)$ for the XY universality class (the explicit expression is
  reported in Ref.~\cite{BPV-21-ccb}; it is valid in the range $[0,
    1.1]$ with an error of at most $0.5\%$).  The inset shows $[U -
    F(R_\xi)]L^{\omega}$ versus $R_\xi$, using the XY
  correction-to-scaling exponent $\omega=0.789$. Data show a
  reasonable scaling behavior as predicted by Eq.~(\ref{udiff}).}
\label{3col2flav-v0-Urxi}
\end{figure}

The FSS analysis of the data at $v=0$ and $\gamma=0$, see
Figs.~\ref{3col2flav_v0} and \ref{3col2flav-v0-Urxi}, provides a clear
evidence of a continuous transition at $\beta_c \approx 4.84$.  If we
plot $R_\xi$ versus $(\beta - \beta_c)L^{1/\nu}$, using the XY
correlation-length exponent~\cite{CHPV-06,Hasenbusch-19,CLLPSSV-19}
$\nu = 0.6717(1)$, we obtain an excellent collapse of the data,
confirming that the transition belongs to the XY universality class.
Fits of $R_\xi$ using the XY estimate for the critical exponent $\nu$
lead to an accurate estimate of the critical point,
$\beta_c=4.8374(2)$. The best evidence for an XY critical behavior is
provided by the plots of $U$ versus $R_\xi$. The data approach the
universal curve of the XY universality class obtained by MC
simulations of the standard XY model.  Differences get smaller and
smaller with increasing $L$. Moreover, see the inset of
Fig.~\ref{3col2flav-v0-Urxi}, deviations are consistent with the
expected FSS scaling behavior
\begin{equation}
  U(L,R_\xi) - F(R_\xi) \approx L^{-\omega} F_{\omega}(R_\xi)\,,
  \label{udiff}
\end{equation}
where $F(R_\xi)$ is the universal curve associated with the XY
universality class, $\omega=0.789(4)$ is the leading XY scaling
correction exponent~\cite{Hasenbusch-19}, and $F_{\omega}(R_\xi)$ is a
scaling function that is universal apart from a multiplicative factor.

\begin{figure}[tbp]  
\includegraphics[width=0.95\columnwidth, clip]{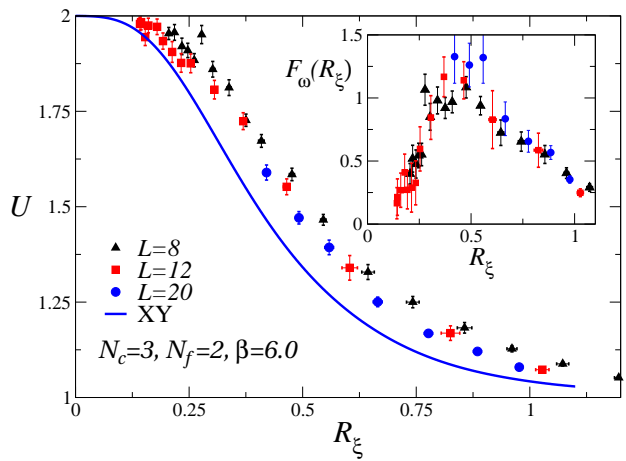}
\caption{Plot of $U$ versus $R_\xi$ for $N_c=3, N_f=2$, $\gamma=0$,
  and $\beta=6.0$. The inset shows $[U - F(R_\xi)]L^{\omega}$ versus
  $R_\xi$, using the XY value $\omega=0.789$, confirming the expected
  behavior (\ref{udiff}). Note also that the data reported in the
  inset appear to collapse onto a curve which differs from that
  reported in the inset of Fig.~\ref{3col2flav-v0-Urxi} only by a
  multiplicative factor, in agreement with Eq.~(\ref{udiff}).}
\label{3col2flav_beta6}
\end{figure}

\begin{figure}[tbp]
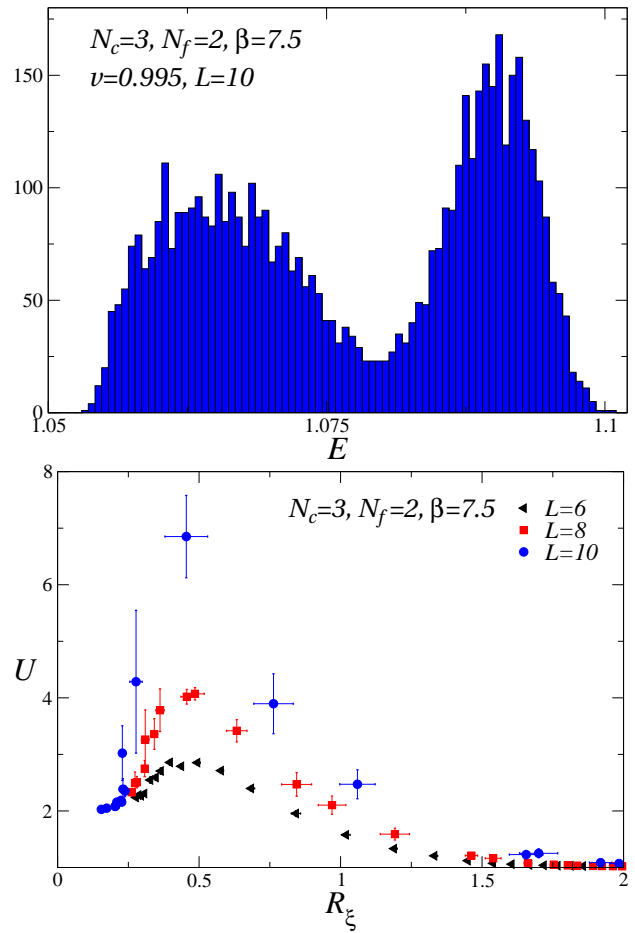
  
\includegraphics[width=0.95\columnwidth, clip]{hist_en_beta7p5.eps}
\includegraphics[width=0.95\columnwidth, clip]{sun3c2fbeta7p5_urxi.eps}
\caption{Top: Histogram of the total energy $E$ for $N_c=3, N_f=2,
  \beta=7.5, v=0.995$, and $L=10$.  Bottom: Binder parameter $U$
  versus $R_\xi$. Data clearly indicate that the transition is of
  first order.}
\label{2col4flav_beta7.5}
\end{figure}

We have also performed simulations at fixed $\beta$, varying $v$. The
numerical results show evidence of an XY continuous transition for
$\beta=5.2$ and $\beta=6$ (see Fig.~\ref{3col2flav_beta6}), at
$v_c\approx 0.23$ and $v_c\approx 0.58$, respectively. On the other
hand, we observe first-order transitions for $\beta=7.5$ at
$v_c\approx 0.99$ (see Fig.~\ref{2col4flav_beta7.5}), for $\beta=9$ at
$v_c\approx 0.75$, and for $\beta=12$ at $v_c\approx 0.45$. Note that
these results are consistent with the fact that the transitions become
of first order as $\beta$ increases and that $v_c\to 0^+$ in the limit
$\beta\to\infty$ (see Sec.~\ref{betainf}).

We do not expect the phase diagram to change for finite $\gamma>0$,
since the main features of the disordered and of the Higgs phase
should not depend on $\gamma$. On the other hand, for $\gamma=\infty$,
the phase diagram should significantly change, see
Sec.~\ref{gammainfty}. One expects three different phases: one
disordered phase, and two different ordered phases, characterized by
different breakings of the global O($N_f$)$\otimes$O($N_c^2-1$) i.e.,
O(2)$\otimes$O(8) in the case at hand (for $\beta \to \infty$, they
would be specified by the sign of $v$). Correspondingly, we expect
three transition lines: one line separates the two ordered phases
(starting at $\beta = \infty$, $v=0$) and two lines separate the
ordered phases from the disordered one.  The order-disorder
transitions for $v>0$ may be continuous, and associated with the
stable fixed point of the corresponding LGW
theory~\cite{PRV-01,PRV-01-ln}.  On the other hand, first-order
transitions are expected for $v<0$, since there is no corresponding
stable fixed point. We do not expect the $\gamma=\infty$ phases to be
stable with respect to the gauge perturbation (this can be proved
using the $\epsilon$ expansion for the simpler case $v=0$), and thus
the $\gamma=\infty$ transitions should only give rise to crossover
effects.

\section{Results for $N_c=2$ and $N_f=4$}
\label{numresnf4nc2}

We now present a study of the phase diagram for $N_f=4$ and $N_c=2$.
In this case, since $N_f>N_c^2-1$, according to the arguments of
Sec.~\ref{gausym}, different Higgs phases caracterized by different
gauge-symmetry patterns are possible. For $\beta\to \infty$, they
correspond to the behavior of the system for $v<0$ and $v>0$, and thus
we will refer to the two phases as the negative-$v$ and positive-$v$
phases, respectively, although, this characterization will not hold
for finite $\beta$. For finite $\beta$ the two phases are divided by a
transition line that ends at $v = 0$, $\beta = \infty$ and which is
expected to be of first order as it is the boundary of two different
ordered phases.

The structure of the negative-$v$ Higgs phase has been discussed in
Sec.~\ref{negvcase}. The global symmetry breaking pattern is
$\hbox{O(4)} \to \hbox{O(3)} \oplus {\mathbb Z}_2$ and the gauge
symmetry breaking pattern is $\hbox{SU(2)} \to \hbox{U(1)}$.  Since
the remnant gauge-invariance group of the Higgs phase is U(1) and a
U(1) gauge theory never undergoes phase transitions, we expect the
gauge coupling to be irrelevant: we have a single negative-$v$ Higgs
phase, irrespective of the value of $\gamma$.  As discussed in
Sec.~\ref{negvcase}, the transition line separating the negative-$v$
Higgs phase from the disordered phase is expected to be described by
the RP$^{3}$ model, which can only undergo first-order transitions.

\begin{figure}[tbp]
\includegraphics[width=0.95\columnwidth, clip]{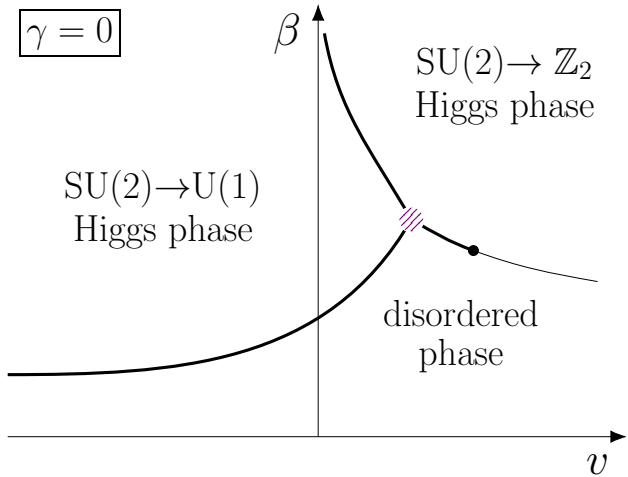}
\caption{Sketch of the phase diagram for $N_c=2$, $N_f=4$, and $\gamma
  = 0$, as inferred from the numerical results. Thick lines denote
  first-order transitions, while the thin line corresponds to
  continuous transitions.  The shaded point is a first-order
  multicritical point (satisfying $1.6 < \beta_{mc} < 2.5$); the
  filled black point that separates first-order from continuous
  transitions occurs at $v=v^*$ with $6 < v^* < 12$.  }
\label{phdianc2nf4}
\end{figure}

The structure of the positive-$v$ Higgs phase is more interesting.
Indeed, the gauge symmetry breaking pattern is $\hbox{SU(2)} \to
{\mathbb Z}_2$, i.e., the Higgs phase is only invariant under the
center of the gauge group. Since ${\mathbb Z}_2$ gauge theories have a
finite-temperature transition, we expect $\gamma$ to be relevant, as
discussed in Sec.~\ref{betainfkappa}. Therefore, we may have two
different positive-$v$ Higgs phases, which differ for the behavior of
the topological modes associated with the gauge-group
center~\cite{SSST-19,SPSS-20}.  The global symmetry breaking pattern
of the positive-$v$ Higgs phase is O(4)$\to$O(3). This would suggest
that the continuous transitions between one of the positive-$v$ Higgs
phases and the disordered phase belong to the O(4) vector universality
class, provided that gauge modes are irrelevant at the transition.

In Fig.~\ref{phdianc2nf4} we report a sketch of the phase diagram for
$\gamma = 0$. As already observed for $N_c=3$ and $N_f=2$, the
negative-$v$ phase extends in the positive-$v$ region for intermediate
values of $\beta$. The transitions between the two low-temperature
phases and between the negative-$v$ and the disordered phases are of
first order, as expected.  The nature of the transition between the
positive-$v$ and the disordered phase depends instead on $v$. At least
for $v\lesssim 6$, the transition line is of first order. On the other
hand, for large $v$, the transitions become apparently continuous.

\begin{figure}[tbp]
\includegraphics[width=0.95\columnwidth,clip]{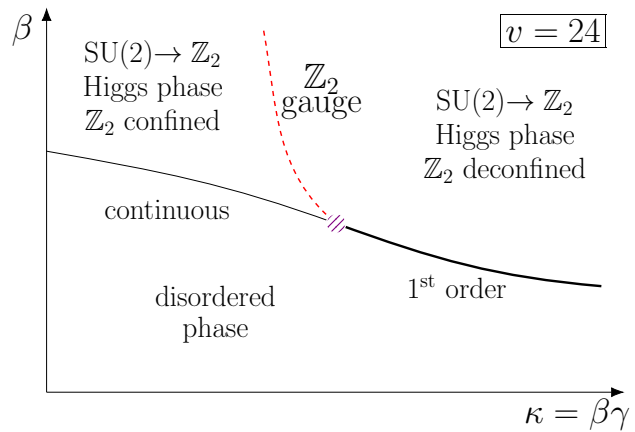}
\caption{A sketch of the $\beta$-$\kappa$ phase diagram for the model
  with $N_c=2$, $N_f=4$ for $v=24$. The ${\mathbb Z}_2$ gauge
  transition line starts at $\kappa_c \approx 0.761$, $\beta =
  \infty$. The multicritical point, where the three transition lines
  meet, satisfies $1 < \kappa < 2$. }
\label{phdiav24nc2nf4}
\end{figure}

To understand the role played by the parameter $\gamma$ for $v > 0$,
we focus on the phase diagram for a specific positive value of $v$, as
a function of $\kappa = \beta \gamma$ and $\beta$. In particular, we
consider the relatively large value $v=24$.  For this value, at
$\gamma=0$, there is a continuous transition between the positive-$v$
Higgs phase and the disorderd phase. A sketch of the phase diagram is
reported in Fig.~\ref{phdiav24nc2nf4}. It is characterized by three
phases, a small-$\beta$ disordered phase, and two large-$\beta$ Higgs
phases, which are distinguished by the behavior of gauge-group center
modes. These phases are separated by three transition lines: (i) a
disordered-Higgs transition line for small $\kappa$, which appears to
be continuous; (ii) a disordered-Higgs transition line for large
$\kappa$, which is of first order; (iii) a continuous ${\mathbb
  Z}_{2}$ gauge (Ising) transition line, which separates the two
low-temperature Higgs phases.

\subsection{The case $\gamma=0$}
\label{gamma0nc2nf4}

\begin{figure}[tbp]  
\includegraphics[width=0.95\columnwidth, clip]{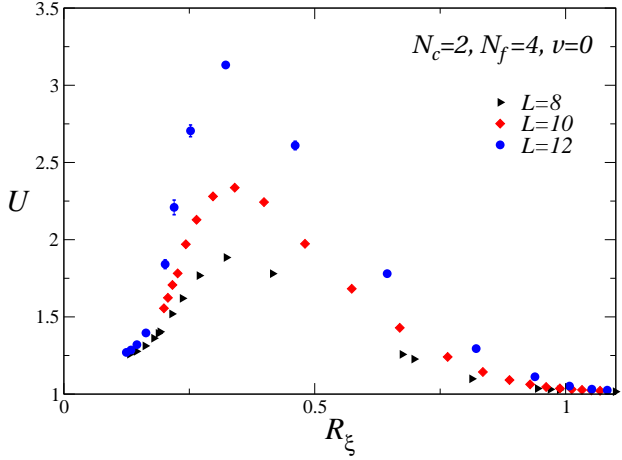}
\caption{Plot of $U$ versus $R_\xi$ for $N_c=2$, $N_f=4$, $\gamma=0$,
  and $v=0$. The rapid increase of the maximum of $U$ indicates that
  the transition is of first order.}
\label{2col4flav_v0}
\end{figure}

\begin{figure}[tbp]
  \includegraphics[width=0.95\columnwidth, clip]{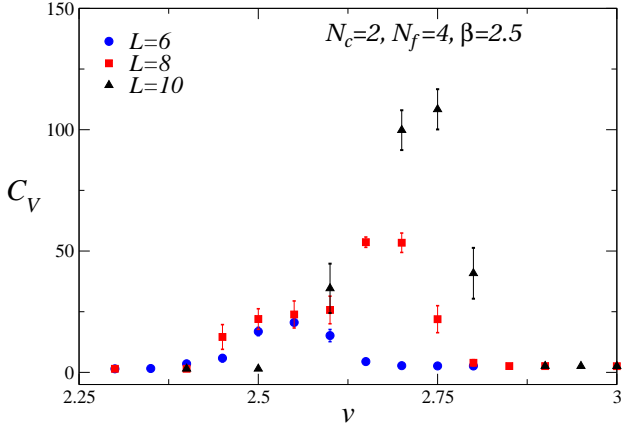}
  \caption{The specific heat, defined in Eq.~(\ref{ecvdef}), versus
    $v$, for $N_c=2, N_f=4$, $\gamma=0$, and $\beta=2.5$, The data
    provide evidence of a first-order transition for $v\approx 2.7$.
  }
\label{sun2c4fbeta2.5}
\end{figure}

To verify that the line that separates the negative-$v$ Higgs phase
from the disordered phase is of first order, we have studied the model
for $v=0$. A transition is observed for $\beta\approx 1.63$. Since the
Binder parameter $U$, reported as a function of $R_\xi$ in
Fig.\ref{2col4flav_v0}, has a maximum that increases rapidly with the
size of the lattice, we conclude that the transition is of first
order.  To verify that transitions along the line that separates the
two Higgs phases are of first order, we have performed simulations at
fixed $\beta=2.5$.  We observe a transition for $v\approx 2.7$. On
both sides of the transition, the Binder parameter $U$ is
approximately 1, as expected, while it increases rapidly for $v\approx
2.7$. The transition is of first order as also confirmed by the
behavior of the specific heat $C_V$, that appears to diverge roughly
as the volume $L^3$, see Fig.~\ref{sun2c4fbeta2.5}.

\begin{figure}[tbp]
  \includegraphics[width=0.95\columnwidth, clip]{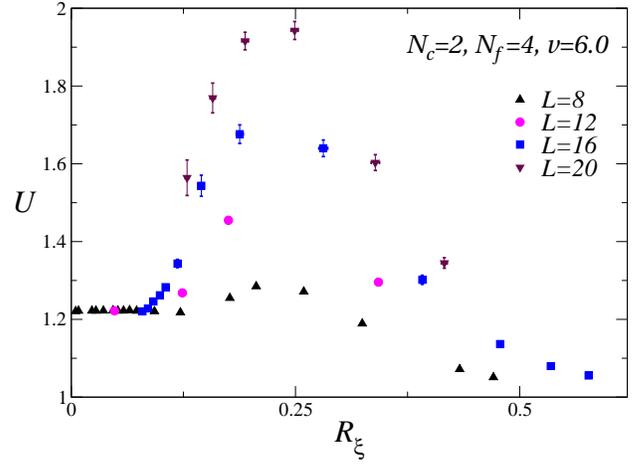}
  \caption{Plot of $U$ versus $R_\xi$ for $N_c=2$, $N_f=4$,
    $\gamma=0$, and $v=6$. The increase of the maximum of $U$ may be
    considered as an early indication of a first-order phase
    transition.}
\label{uvsrxinc2nf4v6}
\end{figure}

\begin{figure}[tbp]
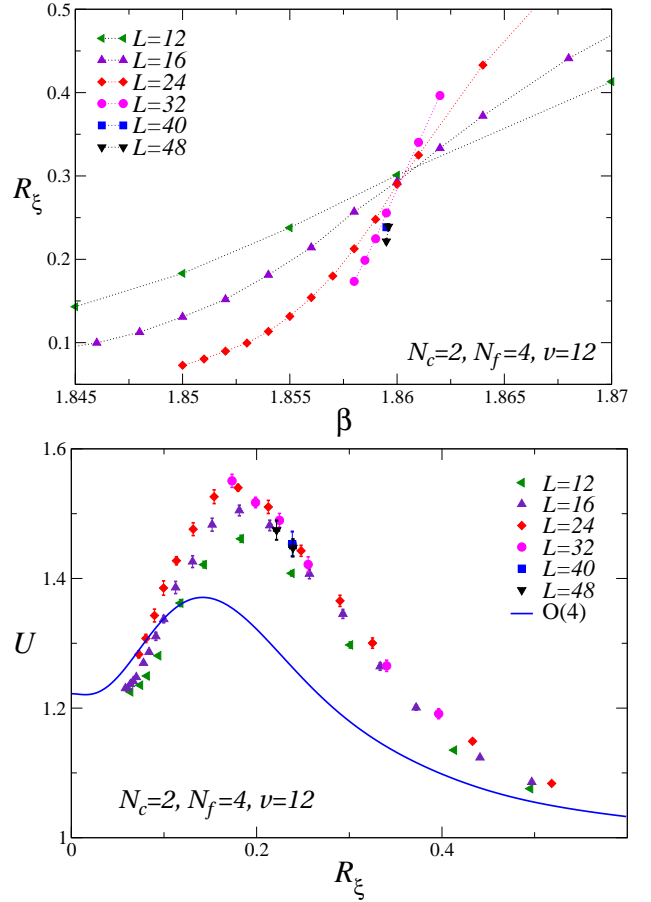

  \includegraphics[width=0.95\columnwidth,
    clip]{sun2c4fv12_rxi_beta.eps}
  \includegraphics[width=0.95\columnwidth, clip]{sun2c4fv12_urxi.eps}
  \caption{Data for $N_c=2, N_f=4$, $\gamma=0$ and the values $v=12$. 
    For comparison, we also report the spin-2 universal curve computed 
    in the O(4) vector model \cite{footnote-O4curve}.
    }
\label{uvsrxinc2nf4vge12}
\end{figure}

We now focus on the transition line separating the disordered phase
from the positive-$v$ Higgs phase, performing simulations at fixed $v$
(we consider $v=6,12,24$, and 48). For $v = 6$, we have studied the
behavior of the system for $0\le \beta \le 3.4$, identifying a single
transition for $\beta_c\approx 2.04$. This guarantees us that the
transition point belongs to the positive-$v$ transition line. The
transition appears to be of first order.  Indeed, the Binder parameter
$U$ has a maximum that increases with increasing lattice size, see
Fig.~\ref{uvsrxinc2nf4v6}. As already mentioned, this behavior
provides an early indication for a first-order transition. Indeed, at
a continuous transition the maximum of $U$ does not increase.

\begin{figure}
  \includegraphics[width=0.95\columnwidth, clip]{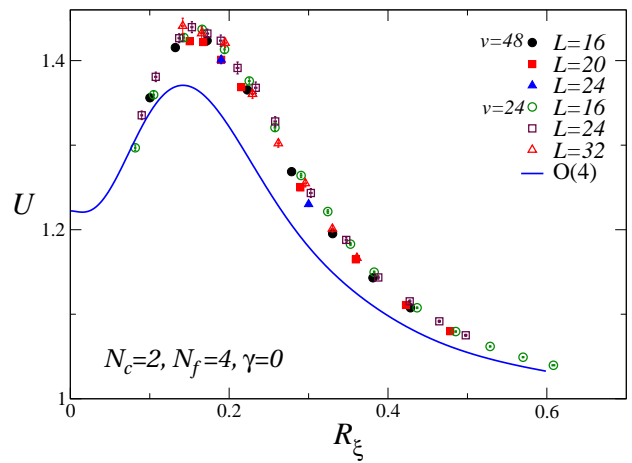}
  \caption{Data for $N_c=2, N_f=4$, $\gamma=0$ and the values $v=24$
    and $v=48$.  For comparison, we also report the spin-2 universal
    curve computed in the O(4) vector model \cite{footnote-O4curve}.}
\label{uvsrxinc2nf4vge24and48}
\end{figure}

The results for $v=12$ (up to $L=48$), $v=24$ (up to $L=32$), and
$v=48$ (up to $L=24$) are consistent with continuous transitions, see
Figs.~\ref{uvsrxinc2nf4vge12} and \ref{uvsrxinc2nf4vge24and48},
located at $\beta_c\approx 1.860$, $\beta_c\approx 1.710$, and
$\beta_c\approx 1.618$, respectively. Fits of $R_\xi$ allow us to
estimate $\nu\approx 0.7$ in all cases, which is quite different from
the effective exponent $\nu=1/d\approx 0.33$ expected at first-order
transitions. These results confirm the phase diagram reported in
Fig.~\ref{phdianc2nf4}: The positive-$v$ transition line is of first
order from the multicritical point, where the three transition lines
meet, up to a tricritical point $v^*$ (with $6\le v^*\le 12$), and
continuous for $v > v^*$.

As mentioned in Sec.~\ref{posvcase}, along the transition line
dividing the disordered phase from the positive-$v$ Higgs phase, the
global symmetry breaking pattern is O(4)$\to$O(3), which is the one
characterizing the vector O(4) universality class.  One would thus
expect O(4) transitions for all $v > v^*$. Fits of $R_\xi$ give
$\nu\approx 0.7$ (with somewhat large errors), which is consistent
with the O(4) value
$\nu=0.750(2)$~\cite{HV-11,Deng-06,Hasenbusch-01,PV-02}.  However, the
scaling curves of the Binder parameter $U$ versus $R_\xi$ are
significantly different from the O(4) one, see
Figs.~\ref{uvsrxinc2nf4vge12} and~\ref{uvsrxinc2nf4vge24and48}.  O(4)
behavior is apparently possible only if there are slowly-decaying and
nonmonotonic scaling corrections. Alternatively, it is possible that
the transitions belong to a new universality class. However, one
should still explain the significant differences in the behavior of
$U$ versus $R_\xi$ for $v=12$ and for $v=24$, 48 (compare
Figs.~\ref{uvsrxinc2nf4vge12} and
\ref{uvsrxinc2nf4vge24and48}). Scaling nonuniversal corrections or
crossover effects due to the nearby tricritical point may be invoked
as possible reasons. In this scenario, the O(4) fixed point would be
unstable, and it would only give rise to crossover phenomena, that
apparently become less important as $v$ increases.  The available
simulations do not allow us to clarify this point. Simulations on
significantly larger lattices are clearly required.

\subsection{The case $\gamma>0$}
\label{gammafnc2nf4}

We have also performed a numerical study of the model for $\gamma>0$,
focusing on the region $v>0$, where gauge-center modes can give rise
to finite-$\gamma$ transitions. We have fixed $v = 24$, obtaining the
phase diagram reported in Fig.~\ref{phdiav24nc2nf4}. We parametrize
the phase diagram in terms of $\kappa\equiv \beta\gamma$ instead of
$\gamma$, since this is the natural variable that appears in the
${\mathbb Z}_2$ gauge model obtained for large values of $\beta$, see
Eq.~(\ref{zncgauge}).  To identify the nature of the disorder-Higgs
transitions, we have performed simulations keeping $\kappa$ fixed and
varying $\beta$. Since the ${\mathbb Z}_2$ gauge transition line ends
at $\kappa_c = 0.761$, $\beta = \infty$, we expect the multicritical
point to have $\kappa_{mc}$ of order 1, and therefore we have
considered $\kappa=1\,,2\,,3$.  Finally, we have performed a
simulation keeping $\beta$ fixed ($\beta = 1.7$) and varying $\kappa$,
to determine the position of the ${\mathbb Z}_2$ gauge transition line
and the corresponding universality class.

\begin{figure}[tbp]
  \includegraphics[width=0.95\columnwidth, clip]{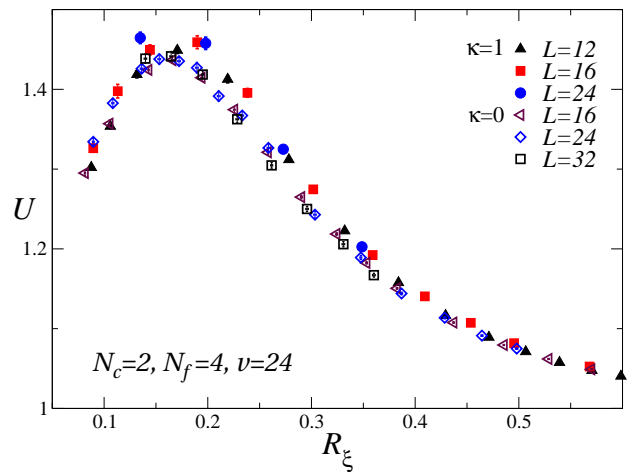}
  \caption{Data of $U$ versus $R_\xi$ for $N_c=2, N_f=4$, $v=24$, and two
   values of $\kappa$, $\kappa=0$ and 1.  }
\label{uvsrxinc2nf4v24gamma1}
\end{figure}

\begin{figure}[tbp]
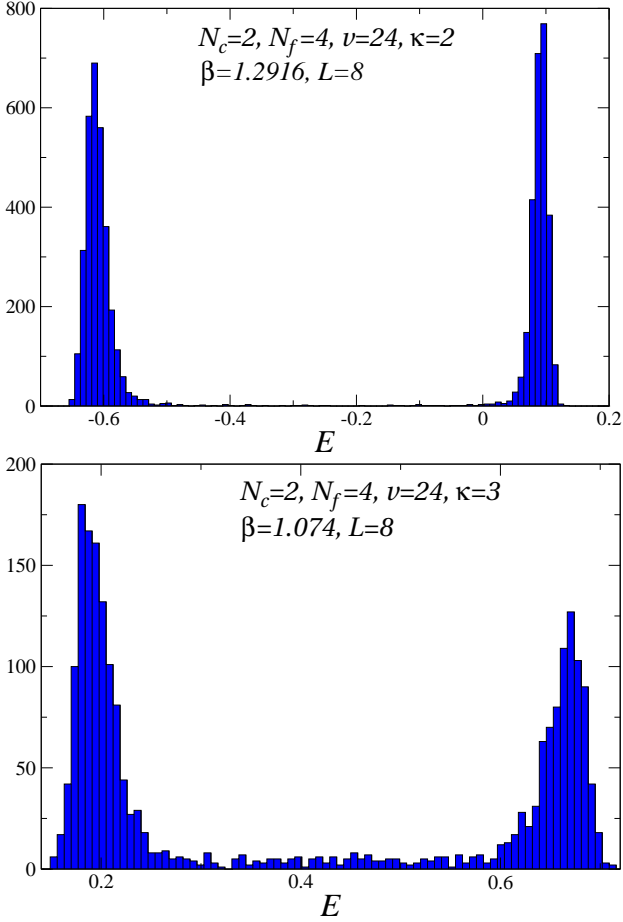

\includegraphics[width=0.95\columnwidth,clip]{hist_entot_k2_beta1p2916_8L.eps}
\includegraphics[width=0.95\columnwidth,clip]{hist_entot_k3_beta1p074_8L.eps}
\caption{Energy hystograms for $v=24$, $\kappa=2$ (top) and $\kappa=3$
  (bottom), for $L=8$. They clearly show the double-peak structure
  characterizing first-order transitions.  The difference of the
  energies of the two maxima provides the latent heat:
  $\Delta_h\approx 0.7$ for $\kappa=2$, and $\Delta_h\approx 0.5$ for
  $\kappa=3$.}
\label{enehystka23}
\end{figure}

For $\kappa=1$, there is a clear evidence of a continuous transition
at $\beta_c\approx 1.615$ (correspondingly $\gamma\approx 0.62$).  The
transition appears to be analogous to that observed for
$\gamma=\kappa=0$ at a similar value of $\beta$ ($\beta_c\approx
1.710$).  In Fig.~\ref{uvsrxinc2nf4v24gamma1} we report $U$ versus
$R_\xi$ for $\kappa = 0$ and 1. Data are consistent with a single
asymptotic curve, suggesting that the two transitions belong to the
same universality class.  Differences are small, of the same order of
the differences observed for the largest $v$ results for $\gamma=0$,
and can be interpreted as scaling corrections.

For $\kappa = 2$ and 3 we observe instead strong first-order
transitions.  For example, the energy distributions are bimodal for
$\kappa=2$, $\beta_c\approx 1.29$ and for $\kappa=3$, $\beta_c\approx
1.07$ (correspondingly $\gamma\approx 1.55$ and $\gamma\approx 2.79$)
already for lattices sizes $L=6,\,8$, see Fig.~\ref{enehystka23}.  The
latent heat $\Delta_h$ is quite large. It decreases with increasing
$\kappa$, varying from $\Delta_h\approx 0.7$ at $\kappa=2$ to
$\Delta_h\approx 0.5$ at $\kappa=3$. This decrease is also confirmed
by results for $\kappa=5$: for $L=8$ the energy distribution is
broad---therefore it is consistent with a first-order transition---but
it does not yet show two peaks.  We have also performed some
simulations for $\kappa=\gamma = \infty$, i.e., of the scalar model
with global O(4)$\otimes$O(3) symmetry, to determine the critical
behavior of the endpoint of the finite-$\kappa$ transition line.  MC
data for relatively small lattices, up to $L=18$, (not shown) are
compatible with a continuous transition (larger lattices are however
needed to confirm this behavior), indicating that $\Delta_h\to 0$ as
$\kappa \to \infty$.

The above-reported results show that the nature of the transition
changes significantly with increasing $\kappa$. While a continuous
transition occurs for $\kappa\lesssim 1$, for $\kappa\ge 2$
transitions are of first order, decreasing their strength with
increasing $\kappa$. A natural hypothesis is that this abrupt change
is due to the different nature of the Higgs phase: For $\kappa = 1$
the low-temperature phase is characterized by confined ${\mathbb
  Z}_{2}$ gauge excitations, while for $\kappa \ge 2$ the ${\mathbb
  Z}_{2}$ gauge modes are deconfined. This requires the existence of
the ${\mathbb Z}_2$ gauge transition line and implies that, in the
sketch reported in Fig.~\ref{phdiav24nc2nf4}, the multicritical point
lies in the region $1 < \kappa_{mc} < 2$.

\begin{figure}[tbp]
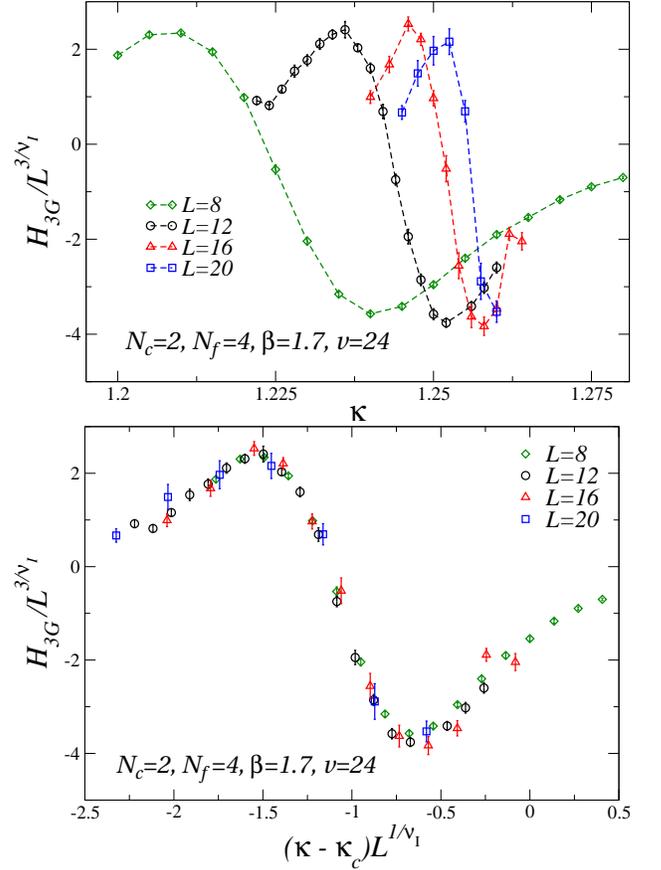

\includegraphics[width=0.95\columnwidth,clip]{momentoterzogauge_vs_k_beta1p7.eps}
\includegraphics[width=0.95\columnwidth,clip]{x3_scaling_beta1p7.eps}
\caption{Estimates of $H_{3G} L^{-3/\nu_I}$ for $v=24$, $\beta=1.7$,
  $N_c=2$, and $N_f=4$, using the Ising exponent $\nu_I = 0.629971$.
  The top panel reports $H_{3G} L^{-3/\nu_I}$ versus $\kappa$, while
  the bottom panel reports $L^{-3/\nu} H_{3G}$ versus
  $L^{1/\nu}(\kappa-\kappa_c)$, with $\kappa_c=1.265$. The plots
  provide evidence of an Ising transition at $\beta=1.7$ and
  $\kappa_c\approx 1.265$. }
\label{H3g-beta1p7}
\end{figure}

We performed simulations to identify the ${\mathbb Z}_2$ gauge
transition line.  We fixed $\beta=1.7$ (which is slightly larger than
the critical point $\beta_c=1.615$ for $\kappa=1$) and varied $\kappa$
between $\kappa=1$ and $\kappa=2$.  To determine the ${\mathbb Z}_2$
gauge transition we monitored thermodynamic quantities, since the
transition is not characterized by a local order parameter.  We
considered cumulants of the gauge part $H_G$ of the Hamiltonian,
focusing on the second and third cumulant. The second cumulant per
unit volume behaves as the specific heat $C_V\sim c\,L^{\alpha/\nu} +
C_{\rm reg}$, where $C_{\rm reg}$ is the regular contribution. Using
the accurate estimates of the 3D Ising critical exponents~\cite{GZ-98,
  CPRV-02, Hasenbusch-10, KPSV-16, KP-17, Hasenbusch-21}, and in
particular~\cite{KPSV-16} $\nu=0.629971(4)$, we obtain
$\alpha/\nu=2/\nu-3 = 0.17475(2)$. The divergence is very mild and
scaling corrections (due to the regular background) decay only as
$L^{-0.17}$, so that FSS analyses of this quantity are not useful for
accurate checks of the Ising behavior. A more promising quantity is
the third cumulant of $H_G$
\begin{equation}
\label{H3G}
H_{3G}=-\frac{1}{\gamma^3}\langle (H_G-\langle H_G\rangle)^3\rangle 
\,.
\end{equation}
It behaves as $H_{3G}\sim L^{3/\nu} = L^{4.76}$, with scaling
corrections that decay as $L^{-\omega} \sim L^{-0.8}$, therefore
significantly faster than in the second-cumulant case.  We will use
the third cumulant to verify the Ising nature of the transition,
checking that the data of $L^{-3/\nu}H_{3G}$ asymptotically collapse
onto a universal scaling function with a peculiar oscillating
shape~\cite{SSNHS-03}, when they are plotted against
$(\kappa-\kappa_c)L^{1/\nu}$, where $\kappa_c$ is the critical point
at $\beta=1.7$. This is nicely confirmed by Fig.~\ref{H3g-beta1p7}. We
have therefore a robust evidence of an Ising transition at
$\kappa_c\approx 1.265$, (correspondingly
$\gamma_c=\kappa_c/\beta\approx 0.744$).

These results provide evidence for the existence of a finite-$\beta$
${\mathbb Z}_2$ transition line, starting at $\kappa_c \approx 0.761$,
$\beta = \infty$, consistently with the sketch reported in
Fig.~\ref{phdiav24nc2nf4}.  Moreover, they suggest that the
multicritical point lies in the region $1 \le \kappa_m \le 2$,
explaining the different behavior observed for $\kappa=1$ and
$\kappa=2,3$.

\section{Conclusions}
\label{conclu}

We have investigated the phase diagram, and the transitions separating
the different phases, of a class of 3D lattice non-Abelian SU($N_c$)
gauge models with $N_f$ ($N_f>1$) degenerate scalar fields in the
adjoint SU($N_c$) representation, using the Wilson formulation of
lattice gauge theories, see Eq.~(\ref{hgauge}).  These models are also
relevant phenomenologically, in particular the model with $N_c = 2$
and $N_f=4$ has been recently proposed to describe optimal doping
criticality in cuprate high-$T_c$
superconductors~\cite{SSST-19,SPSS-20}.

We discuss the role played by the scalar quartic potential and by the
gauge-group representation of the scalar fields, which are crucial to
determine the structure of the low-temperature Higgs phases and the
nature of the phase transitions.  For this purpose we have performed a
detailed analysis of the minima of the scalar-field potential.  As
discussed in Sec.~\ref{gausym}, such an analysis shows the emergence
of two qualitatively different phase diagrams, depending on the number
of colors $N_c$ and of flavors $N_f$.  For $N_f\le N_c^2-1$ a single
Higgs phase exists and, for positive values of $v$ ($v$ is parameter
entering the scalar potential (\ref{potential})), there is a single
disordered phase for any temperature up to $T=0$. For $N_f>N_c^2-1$,
instead, two different low-temperature Higgs phases exist, with
transitions characterized by different global and gauge symmetry
breaking patterns.  In particular, for $N_c=2$, we have a
low-temperature Higgs phase characterized by the gauge symmetry
breaking pattern ${\rm SU}(2) \to {\rm U}(1)$ (this phase is observed
when $v$ is negative), and a second low-temperature Higgs phase
characterized by ${\rm SU}(2) \rightarrow {\mathbb Z}_{2}$ (this
occurs for positive $v$).

The phase diagram of the model can also be influenced by the
properties of the gauge modes, depending on the residual gauge
symmetry present in the ordered Higgs phase.  The phase diagram for
$N_f\le N_c^2-1$ is not expected to depend on the gauge modes, as the
residual gauge symmetry group in the Higgs phase is continuous, and
therefore no finite-temperature transitions associated with these
gauge variables are possible in three dimensions.  Thus, the phase
diagram should not depend on the gauge coupling $\gamma$.  Differences
should occur only for $\gamma=\infty$. Analogously, for $N_f >
N_c^2-1$, the negative-$v$ Higgs phase should not depend on $\gamma$,
given the large residual gauge symmetry characterizing its
low-temperature Higgs phase.  On the other hand, in the positive-$v$
Higgs phase present for $N_f>N_c^2-1$ configurations are only
invariant under gauge-group center transformations.  Since ${\mathbb
  Z}_{N_c}$ gauge theories undergo finite-temperature transitions,
there are two different Higgs phases, characterized by the same
gauge-symmetry breaking pattern SU($N_c$)$\to {\mathbb Z}_{N_c}$, but
differing in the topological behavior of the ${\mathbb Z}_{N_c}$ gauge
modes. as sketched in Fig.~\ref{phdiav24nc2nf4} for $N_c=2$ and
$N_f=4$.

We have presented numerical studies of two representative models: the
case $N_c=3$, $N_f = 2$, to verify the general scenario for models
satisfying $N_f \le N_c^2 - 1$, and the case $N_c=2$, $N_f = 4$, which
shows the more complex phase diagram predicted for models satisfying
$N_f > N_c^2 - 1$, and which is also relevant for cuprate
superconductors.  In both cases, the general predictions for the Higgs
phases, and for the nature of the transition lines, are verified.

Although our results confirm the general picture, there are still some
issues that call for further investigations. For $N_f=4$ and $N_c=2$,
we have evidence of continuous transitions for small values of
$\gamma$ and large positive values of $v$, whose characterization is
not clear (we have not been able to assign these transitions to a
known universality class).  A second issue is the behavior for large
values of $\gamma$ and of $v$. We have observed first-order
transitions, whose latent heat decreases with increasing $\gamma$. It
would be interesting to investigate the nature of the endpoint of the
transition line at $\gamma=\infty$: numerical simulations on small
lattices are consistent with a continuous transition, but larger
lattices are needed to settle the question.

An intriguing possibility is that the continuous transitions observed
when $N_f>N_c^2-1$, $v>0$ and small values of $\gamma$, are associated
with the fixed point found in the analysis of the one-loop
$\epsilon$-expansion RG flow, see Sec.\ref{sft}.  The $O(\epsilon)$
fixed point in Eq.~(\ref{fpln}) is stable only if $N_f > N^* \approx
210$ close to four dimensions. However, it is conceivable that the
critical number $N^*$ is drastically smaller in three dimensions, so
small to include $N_f=4$.  One might find this possibility
unplausible; however, we should note that this is what happens in the
Abelian-Higgs U(1) field theory.  A leading order $\epsilon$-expansion
computation~\cite{HLM-74}, analogous to the one reported in
Sec.~\ref{sft}, predicts the existence of a stable fixed point for
$N_f>N_f^*\approx 183$. However, if higher-order corrections are
included \cite{IZMHS-19}, a significantly smaller estimate of the 3D
critical value is obtained.  A numerical MC study in three dimensions
finds $N_f^*=7(2)$~\cite{BPV-21-nc}, confirming that the one-loop
estimate is of no quantitative relevance.  We believe that further
work is called for to test this possibility and to achieve a full
understanding of the actual behavior of the model for positive values
of $v$.  A promising strategy consists in studying the model for
$v=+\infty$. The model is significantly simpler, since the scalar
field variables take the form given in Eq.~(\ref{locpotential}), and
simulations faster, as there is no need to take the potential into
account in the update and more powerful MC algorithms can be used.
This should allow us to perform a more effective numerical study of
the phase diagram in the $\beta$-$\gamma$ plane.

\bigskip

\emph{Acknowledgement}.  Numerical simulations have been performed on
the CSN4 cluster of the Scientific Computing Center at INFN-PISA.

\appendix

\section{Monte Carlo simulations}
\label{mcsim}

We performed MC simulations on cubic lattices with periodic boundary
conditions.  The gauge link variables $U_{{\bm x},\mu}$ were updated using a
standard Metropolis algorithm \cite{Metropolis:1953am}.  The new link variable
was chosen close to the old one, in order to guarantee an acceptance rate of
approximately 30\%. The scalar fields were updated using two different
Metropolis updates, again tuning the proposal to obtain an acceptance rate of
30\%. The first move performs a rotation in flavor space, while the second one
rotates the color components of a single flavor. This update procedure is the
same already used in \cite{BFPV-21}, to which we refer for some more
implementation details.  For the largest sizes simulated, the typical
statistics are of the order of $10^{6-7}$ ($N_c=3, N_f=2$) and of $10^{7-8}$
($N_c=2, N_f=4$) lattice sweeps of both scalar and gauge variables.  To take
into account autocorrelations and determine the correct statistical errors, we
used a standard blocking and jackknife procedure. Our maximum blocking sizes
were of the order of $10^{4-5}$ ($N_c = 3, N_f = 2$) and $10^{5-6}$ ($N_c=2,
N_f=4$).


\begin{thebibliography}{99}
\bibitem{footnote-O4curve} We report here the universal scaling 
curve $U = F(R_\xi)$ computed in the O(4) model. The Binder parameter 
and the correlation length are defined as 
in Sec.~\ref{obsfss}, using $Q^{ab}= s^a s^b - {1\over 4} \delta^{ab}$
($s^a$ is a 4-dimensional unit vector). The parametrization 
$F(x) = 1.22222 - (282926.32646 + 1135938.16294 x^2) x^2 e^{-2 x} +
(35362.99172 + 47446.25649 x) x (1 - e^{-8 x})   +
187496.45984 x^3 + 184283.68973 x^4 -  536152.54309 x^5 +
358418.18420 x^6 - 84956.08026 x^7$
is valid for $x<0.6$ with a relative error of less than $0.5\%$.
\end{thebibliography}
\end{document}